\documentclass[prl,twocolumn,amsmath,amssymb,floatfix]{revtex4}
\usepackage{amsmath}
\usepackage{graphicx}
\usepackage{epsfig}

\usepackage{bm}

\newcommand{\be}{\begin{equation}}
\newcommand{\ee}{\end{equation}}

\newcommand{\beq}{\begin{eqnarray}}
\newcommand{\eeq}{\end{eqnarray}}

\def\H1{\widehat{H}_1}

\DeclareMathOperator{\Tr}{Tr}
\newcommand{\ket}[1]{\left|#1\right\rangle}
\newcommand{\bra}[1]{\left\langle#1\right|}

\newcommand{\braopket}[3]{\left\langle#1\middle| #2 \middle| #3\right\rangle}
\newcommand{\av}[1]{\left\langle#1\right\rangle}
\newcommand{\co}[2]{{#1^\dagger_{#2}}}
\renewcommand{\ao}[2]{{#1^{\vphantom\dagger}_{#2}}}

\def\op#1{{\Hat{\mathrm{#1}}}}

\begin{document}

\title{Dynamical symmetry approach and topological field theory for path integrals of quantum spin systems}

\author{Matou\v{s} Ringel, Vladimir Gritsev}
\affiliation{Physics Department, University of Fribourg, Chemin du
Mus\'{e}e 3, 1700 Fribourg, Switzerland}

\begin{abstract}
We develop a dynamical symmetry approach to path integrals for general interacting quantum spin
systems. The time-ordered exponential obtained after the Hubbard-Stratonovich
transformation can be
disentangled into the product of a finite number of the usual exponentials. This procedure leads to 
a set of stochastic differential equations on the group manifold, which
can be further formulated in terms of
the supersymmetric effective action. This action has the form of the Witten topological field theory
in the continuum limit.
As a consequence, we show how it can be used to obtain the exact results for a specific quantum
many-body system which can be otherwise solved only by the Bethe ansatz. To our knowledge this
represents the
first example of a many-body system treated exactly using the path integral formulation. 
Moreover, our method can deal with time-dependent parameters, which we demonstrate explicitly.
\end{abstract}

\maketitle

\section{Introduction}
The path integral approach to strongly-correlated systems is a powerful method from many perspectives.
In particular, it easily accounts for the topologically nontrivial terms in the action and it is
a convenient starting point for various numerical  schemes. Moreover, it allows to treat
the correlated systems by a number of approximate analytical techniques, including the saddle point
method, instanton analysis, and various perturbative expansions~\cite{ref:Kleinert,ref:Negele,ZJ}. 
On the other hand, only a limited number of path integrals are accessible to 
an exact evaluation, and to our knowledge there are no examples  
that could explicitly treat any nontrivial Bethe ansatz-solvable model. 
For spin systems the conventional method
consists of inserting the resolution of identity on the space of
the spin coherent states at every discretized time slice~\cite{ref:Fradkin}.
Then the overlaps between different coherent states taken at consecutive  time slices
$\langle \zeta(t)|\zeta(t+\epsilon)\rangle$, where $\epsilon$ is the discretization time step, is
approximated using the Taylor expansion in~$\epsilon$. 
The important assumption behind this is the differentiability of
the path $\zeta(t)$. 
This approximation eventually prohibits exact evaluation of the path integral. 

Here we introduce a novel representation of  lattice spin models,
which does not rely on the spin coherent state representation, 
and which reveals a hidden dynamical (super)-symmetry structure. After performing
the Hubbard-Stratonovich transform, the partition function or the evolution operator of a spin
system quadratic in spin variables can be represented as an average of a time-ordered
exponential.  Based on the well known facts of the group theory, we represent the time-ordered
exponential as a product of usual exponentials. 
However, the arguments of the disentangled exponentials
are related to the original fields via a set of nonlinear 
differential equations.  These equations can be interpreted as stochastic differential
equations. The Hubbard-Stratonovich fields play the role of a noise, 
whose correlators 
are defined by the interaction matrix of the ordinal quantum spin system. 
Stochastic trajectories are
non-differentiable in general and our approach takes this into account exactly and allows to derive
results that can be obtained otherwise only using the Bethe ansatz. 
We show this explicitly on a non-trivial example rooted in quantum optics ~\cite{ref:YudsonPRA}. 
Moreover the stochastic interpretation suggests that there is a hidden 
supersymmetry, which eventually leads us to the formulation of the partition
function of the quantum spin system as a correlation function of 
non-topological operators in the
theory whose action is given by the topological field theory of Witten~\cite{Witten,FLN}. 

Sections 2 and 3 of this paper discuss the general aspects of our approach, while Section 4
illustrates the method on a non-trivial example of a many-body system.
We compare our results to the Bethe ansatz solution within the limit of its applicability,
and we also provide some explicit results beyond.
This example provides a hint of utility of our approach for 
a  larger class of  spin systems. The Appendix contains a number of formulas,
useful  for analytical and numerical considerations.

\section{Disentanglement of the time-ordered exponential}
We consider a generic interacting quantum spin model on a lattice,
\beq
H=\sum_{i,j}\Omega_{ij}^{ab}S_{i}^{a}S_{j}^{b}+\sum_{j}h^{a}_{j}S^{a}_{j}
\label{model}
\eeq
where the lattice spin operators $S^{a}_{j}$  ($j$ being the lattice index)
satisfy the commutation relations of the Lie algebra $g$, 
$[S^{a}_{i},S^{b}_{j}]=f^{ab}_{c}S^{c}_{j}\delta_{ij}$. The indices $a,b,c$ run 
from $1$ to $N$, where $N$ is the dimension of $g$, and $f^{ab}_{c}$ are the structure constants. 
Interactions between the spins are determined by the
interaction matrix $\Omega^{ab}_{ij}$, which can be in general non diagonal in $a,b$
indices. We assume no particular restriction on the compactness of the
corresponding Lie group $G$ while we assume that certain particular representation is chosen for the
concrete physical problem (such as the spin-1/2 or spin-1 representations for the case of $g$ being
$SU(2)$). We are interested in  correlation functions of the model defined by Eq.~\eqref{model},
which is why we introduce the source terms~$\mathcal{J}_j^a(t)$ and
consider the generating functional 
\beq
W[{\cal J}]=\Tr\exp\left(it H+i\sum_{j}\int {\cal J}^{a}_{j}(t)S^{a}_{j}\right).
\label{gen-fun}
\eeq
The approach under discussion may work also in a more general 
setting, in which the matrix $\Omega^{ab}$ and the fields $h^{a}$ are
assumed to be time-dependent functions. Moreover, it is straightforward
to replace the trace in Eq.~\eqref{gen-fun}
by expectation values in certain $\ket{\mathrm{in}}$ and $\ket{\mathrm{out}}$. These
generalize the approach to non-equilibrium evolution problems. For the equilibrium
considerations one can perform the Wick rotation $it\rightarrow-\beta$.

Applying the Suzuki-Trotter discretization~\cite{ref:Negele} and introducing the
Hubbard-Stratonovich transformation~\cite{ref:HSTransform}, we rewrite Eq.~\eqref{gen-fun}
into the following form (in the following repeated indices are assumed to be summed over)
\beq
W[{\cal J}]=\int D\mu(\phi)\prod_{j}{\cal T}_{c}\exp(i\int^{t} dt \Phi^{a}_{j}S^{a}_{j}),
\label{W(J)}
\eeq
where $D\mu(\phi)=\exp(-\sum_{jk}\phi^{a}_{k}(\Omega^{-1})^{ab}_{kj}\phi_{j}^{b})\prod_{j}D\phi^{a}$
is the Gaussian integration measure, $\Phi^{a}_{j}(t)=\phi^{a}_{j}(t)+h^{a}_{j}+{\cal J}^{a}_{j}(t)$,
$\Omega^{-1}$ is a matrix inverse to $\Omega$, and ${\cal T}_{c}\exp$ denotes the
time-ordered exponential. The index $c$  indicates a possible Keldysh contour ordering.

The main conceptual step of our approach is the following: the "effective Hamiltonian" $\int dt
\Phi^{a}_j(t)S_j^{a}$ at given $j$ is a linear combination of generators of $g$ and therefore its
exponential is an element of the group $G$. The ${\cal T}$-ordered exponential is a directed
product of ordinary exponentials, and hence a composition of elements of the Lie group. Any
composition of elements of the Lie group~$G$ is a certain element of $G$ itself. As such, this element
can be written as a single exponential again. This generic mathematical fact defines the essence of
our approach and exhibits a concept of dynamical symmetry: A Hamiltonian of the form of a linear
combination of generators of the Lie algebra (with possibly time-dependent coefficients) has this Lie
algebra as the spectrum-generating algebra~\cite{Perelomov}. We call the procedure of going from the
time-ordered exponential to the product of the ordinary exponentials  a \emph{disentanglement
transformation}.
Similar philosophy has been recently used in
\cite{ref:Kolokolov1,ref:Kolokolov2,ref:Kolokolov3,ref:Kolokolov4,ref:Kolokolov5Review,ref:Kolokolov6,Galitski,ref:Gefen1,ref:Gefen2,Chalker,RPG}.

It is clear that the original variables $\Phi^{a}(t)$ and a set of variables, which carries out a
disentanglement are related in a non-trivial way. The easiest way to elucidate this relationship
is  to consider the definition of the ${\cal T}$-ordered exponential, namely
$i\dot{U}(t)=\Phi^{a}(t)S^{a}U$.   Since $U\in G$ and there is an inverse, $U^{-1}$ we write it as
$i\dot{U}U^{-1}=\Phi^{a}(t)S^{a}$.  The left-hand side of this equation is a current $J_{0}$ on the
group, used e.g. for formulating the non-linear sigma models. Group manifold can be parametrized in a number of ways by a
set of parameters $\{n^{a}\}$, ($a=1\ldots \dim G$). However, there is an important object --- the 
Maurer-Cartan (MC) 1-form on $G$ --- that allows us to write the underlying equations in a covariant way.
Defining a right 1-form $R^{a}_{\alpha}$ via
$i\dot{U}U^{-1}=i\partial_{\alpha}UU^{-1}\dot{n}^{\alpha}=iR^{a}_{\alpha}(\{n\})
S_{a}\dot{n}^{\alpha}$, where $n^{\alpha}$ is a parametrization of a group, we arrive at a set of
equations defining a relation between $\Phi^{a}(t)$ and a disentangling variables, 
\beq
iR^{a}_{\alpha}(\{n\})\dot{n}^{\alpha}(t)=\Phi^{a}(t).
\label{MC-eq}
\eeq
For a given parametrization $\{n\}$, the MC forms satisfy the MC equations,
$\partial_{\alpha}R^{a}_{\beta}-\partial_{\beta}R^{a}_{\alpha}+f^{a}_{bc}R^{b}_{\alpha}R^{c}_{\beta}=0$,
which are the defining equations for $R^{b}_{\alpha}$. Here and in Eq.(\ref{MC-eq}) we
for generality distinguish the upper and lower indexes,  which are manipulated with the Killing-Cartan metrics
$g_{\alpha\beta}=R^{a}_{\alpha}R^{b}_{\beta}\Tr(S^{a}S^{b})$.  The equations (\ref{MC-eq}) can in
principle be resolved in terms on $\{\dot{n}^{a}\}$, however this can not be done 
globally if the topology of $G$ is non-trivial (e.g. for the $SU(2)$ case the sphere can not be
covered by a single map). Also note that for $G$ interpreted as the Riemann manifold, MC forms are
proportional to vielbeins (tetrads) $e^{a}_{\beta}$. 

We envision at least two general ways to proceed. First, thanks to the Gaussian measure
$D\mu(\phi)$ of the Hubbard-Stratanovich fields we can interpret Eqs.~(\ref{MC-eq}) as a set of
stochastic differential equations on the group manifold. Second, one can use Eqs.~(\ref{MC-eq}) as 
defining rules for the change of variables in the path integral~(\ref{W(J)}), to go from variables
$\Phi^a$ to the new variables $n^a$ that define a disentanglement transform. Here we elaborate more
on the first approach.

For a concrete implementation of this approach we consider first the case of $g=su(2)$ and come back to
a generic discussion later on. As we have said, many different parametrizations for a group element
$U$ are possible. These include the Euler angle parametrization,
$U_{E}(\alpha,\beta,\gamma)=\exp(i\alpha S^{3})\exp(i\beta S^{2})\exp(i\gamma S^{3})$, the covariant
parametrization, $U_{C}({\bf A})=\exp(i{\bf A}\cdot {\bf S})$, and the Gauss parametrization which can be globally defined on the group manifold \cite{WN}
\begin{equation}
U_{G}(\xi_\pm,\xi_z)=\exp(\xi_+  S^{+})\exp(\xi_z S^{z})\exp(\xi_- S^{-}),
    \label{eq:DisentanglementSU}
\end{equation}
on which we mostly focus.  For the latter parametrization we can derive the following set of equations 
(in imaginary time),
\beq\label{phiABC}
\Phi_{-}&=&\dot{\xi_+}-\xi_+\dot{\xi_z}-\xi_+^{2}e^{-\xi_z}\dot{\xi_-}\nonumber\\
\Phi_{z}&=&\dot{\xi_z}+2\xi_+e^{-\xi_z}\dot{\xi_-}\\
\Phi_{+}&=&e^{-\xi_z}\dot{\xi_-}\nonumber
\eeq
and inversely,
\beq
\dot{\xi_+}&=&\Phi_{-}+\Phi_{z}\xi_+-\Phi_{+}\xi_+^{2},\nonumber\\
\dot{\xi_z}&=&\Phi_{z}-2\Phi_{+}\xi_+,\\
\dot{\xi_-}&=&\Phi_{+}\exp(\xi_z).\nonumber
\label{eq:RiccatiEqsSU}
\eeq
The initial conditions $\xi_{\pm,z}(0)=0$ follow from the relation~$U(0)=1$. For the real time
evolution the factor of $i$ should be present in front of $\dot{\xi}_{\pm,0}$.

The trace of the evolution operator can be easily computed for the spin-1/2 representation of the
evolution operator, $\Tr U_{G}=2\cosh(\xi_z/2)+\xi_+\xi_-\exp(-\xi_z/2)$. For a generic representation
of spin $s$ the covariant form $U_{C}$ yields a compact formula for the characters, $\chi^{s}(|{\bf
A}|)=\sum_{m=-s}^{m=s}\exp(-im|{\bf A}|)=\sin\left[(s+\frac{1}{2})|{\bf
A}|\right]/\sin\left[\frac{1}{2}|{\bf A}|\right]$, while for the $su(N)$ case one can use the
celebrated Weyl determinant formula for the characters \cite{Lie-group,KSK}.

Next, by looking at the Eqs.~(\ref{eq:RiccatiEqsSU}) we realize that only the equation for $\xi_+$ is
independent, while the solutions for $\xi_{z,-}$ can be obtained from the one for~$\xi_+$.
This motivates us to
look for  further change of variables. This can be done in different ways. In particular, by
introducing  new fields $\psi^{\pm,z}$ via the following
correspondence
$\xi_+=\psi^{-}$, $\xi_z=\int_{0}^{t}\psi^{z}(t')dt'$ and
$\xi_-=\int_{0}^{t}\psi^{+}(t')\exp(\int_{0}^{t'} \psi^{z}(t'')dt'')dt'$ we obtain
$\Phi^{z}=\psi^{z}+2\psi^{+}\psi^{-}$, $\Phi^{+}=\psi^{+}$,
and $\Phi^{-}=\dot{\psi}^{-}-\psi^{z}\psi^{-}-\psi^{+}(\psi^{-})^{2}$. This representation was 
found in
\cite{ref:Kolokolov1,ref:Kolokolov2,ref:Kolokolov3,ref:Kolokolov4,ref:Kolokolov6,ref:Kolokolov5Review}
and more recently in \cite{ref:Gefen1,ref:Gefen2}. 
The  measure of integration changes correspondingly,
$D\phi^{z}D\phi^{+}D\phi^{-}={\cal M}D\psi^{z}\psi^{+}D\psi^{-}$, where
${\cal M}_{\psi}=c_{\psi}\exp(-\frac{1}{2}\int_{0}^{t}\psi^{z}(t))
dt)$ is the Jacobian of the transformation.
The constant $c_{\psi}$ fixes the normalization, and can be adjusted e.g. comparing to a
non-interacting system.
The following change of variables 
\beq
\xi_+ &=&\rho^{-}, \qquad \xi_z=\int_{0}^{t}(\rho^{z}-\rho^{+}\rho^{-}),\nonumber\\
\xi_- &=&\int_{0}^{t}\rho^{+}\exp\left[\int_{0}^{t}(\rho^{z}-\rho^{+}\rho^{-})\right],
\eeq
leads to the following equations:
\beq
\Phi^{+} &=&\rho^{+},\nonumber\\
\Phi^{-} &=& \dot{\rho}^{-}-\rho^{-}\rho^{z},\nonumber\\
\Phi^{z} &=&\rho^{z}+\rho^{-}\rho^{+}.
\eeq
The differential equation for $\dot\rho$ can be formally solved,
\beq
\rho^{-}(t)&=&\exp\left(\int_{0}^{t}\rho^{z}(t')dt'\right)\\
&\times&\int_{0}^{t}\exp\left(-\int_{0}^{t'}\rho^{z}(t'')dt''\right)\Phi^{-}(t')dt'\nonumber,
\eeq
where we assumed the initial condition $\rho^{-}(0)=0$. Therefore in this representation we
implicitly resolved the non-linearity of the equations in exchange for the non-locality in time. Variables
$\xi_{\pm,z}$ can be now expressed exactly via $\rho^{z},\phi^{+},\phi^{-}$. Therefore we can use it
to construct some mixture representation which involves $\phi$-variables and $\rho^{z}$
simultaneously.
This representation suggests that the variable
$R(t)=\int_{0}^{t}\rho^{z}(x)dx$ (such that $\dot{R}=\rho^{z}(t)$) considered as an independent
variable might be useful. The Jacobian in this case reads ${\cal
M}_{\rho}=c_{\rho}\exp(\frac{1}{2}\int_{0}^{t}(-\rho^{z}+\rho^{-}\rho^{+}))$. One more example is
given the change of variables
\beq
\xi_+ &=&\lambda^{-}, \nonumber\\
\xi_z &=& \int_{0}^{t}(\lambda^{z}-2\lambda^{+}\lambda^{-}),\\
\xi_- &=&
\int_{0}^{t}\lambda^{+}\exp\left[\int_{0}^{t}(\lambda^{z}-2\lambda^{+}\lambda^{-})\right]\nonumber,
\eeq
which leads to the following equations
\beq
\Phi^{+} &=&\lambda^{+},\nonumber\\ 
\Phi^{-} &=& \dot{\lambda}^{-}-\lambda^{-}\lambda^{z}+(\lambda^{-})^{2}\lambda^{+},\\
\Phi^{z} &=&\lambda^{z}\nonumber
\eeq
The advantage of this representation lies in the simplicity of the relationship between the  $+$ and
$z$-components of the fields. This can be useful for some particular physical situations. Jacobian
of the change of variables from $\phi^{\pm,z}$ to $\lambda^{\pm,z}$ is ${\cal
M}_{\lambda}=c_{\lambda}\exp\left(\frac{1}{2}\int_{0}^{t}(-\lambda^{z}+2\lambda^{-}\lambda^{+})\right)$. 

We note that there is an infinite number of other parametrizations, which could be constructed in a
similar fashion. Based on the above parametrizations, one could conjecture that the determinant of the
transformation in the Gauss decomposition can be expressed in the form of a group element
corresponding to the time averaged Cartan subalgebra.

While in general the Riccati equations encountered above can not be solved analytically, we note
that one can establish an interesting connection to the theory of the KdV (Korteweg-de~Vries)
equation~\cite{ref:KDV}.  Namely, using the projective linear (M\"{o}bius) transformation, the
Riccati equation can always be put into the in the form $iy'(x)+y^{2}(x)=k^{2}-u(x)$ where
$u(x)\equiv u({\bf\Phi}(x))$ is a known function of the HS variables ${\bf\Phi}(x)$,  $x\equiv
x(t)$ being a reparametrized time variable, and $k$ a real parameter (which could related to real
physical parameters, e.g. magnetic field or $\hbar$ in the WKB series). The function $y(x)$ then
admits 
a series expansion $y(x,k)=k+\sum_{n\geq 1}y_{n}(x)(2k)^{-n}$. Here, all even coefficients $y_{2n}$
are  total derivatives and imaginary, while the first few nontrivial terms are $y_{1}=u(x)$,
$y_{3}(x)=u^{2}$, $y_{5}=u^{3}+(u')^{2}/2$, $y_{7}=5u^{4}/2 -5u^{2}u''/2 +(u'' )^{2}/2$, etc.
The integrals $\int dx y_{2n+1}$ form the (Poisson) commuting integrals of the KdV equation. It would be
interesting to understand the implications of this observation on our path-integral formulation
below. 

It can be shown that a generalization of the disentangling transformation for other groups --- such as
$SU(N)$ ---  is straightforward: The resulting set of equations will have the form of the matrix Riccati
equation. On the other hand, a generalization for  super-groups is more tricky and should be
studied case by case, especially for  atypical representations.

\section{Stochastic interpretation, supersymmetry and topological field theory formulation} 
The Hubbard-Stratonovich transformation introduces a Gaussian measure for the fluctuating fields
$\phi^{a}_{j}$. This suggests that the differential equations for the disentangling variables
$\xi_{\pm},x_{z}$ can be interpreted as a set of stochastic differential (generalized Langevin)
equations. Generically, solutions of these equations are non-differentiable paths which should be
treated carefully using the appropriate discretization prescriptions (see Appendix). This leads to
the effective Lagrangian formulation (path integral) formulation and, on the other hand, reveals the
hidden super-symmetry structure of the lattice spin systems. Here we establish the connection between
our disentanglement procedure and the theory of stochastic differential equations. 

\subsection{Stochastic interpretation}
Provided the MC form (vielbein) is invertible, the disentangling equations can be put into the form
\beq 
i\dot{n}^{\alpha}_{j}=(R^{a}_{\alpha})^{-1}_{j} \Phi^{a}_{j} 
\label{eq:stoch1} 
\eeq 
for every
lattice site $j$. The path integral formulation of the stochastic processes is a delicate issue
and has a long history. The difficulty comes from the proper definition of the continuous-in-time
effective Lagrangian from the discretized version of the stochastic processes. This issue was
carefully elaborated upon in a series of earlier works \cite{LRT} (and rediscovered recently in \cite{Arnold}) where it was shown that the result for
the effective Lagrangian is a two-parametric family labeled by $r,s\in[0,1]$ which stand for
discretization of the stochastic variable and the noise respectively. While the mid-point
discretization $r=s=1/2$ corresponds to the Stratonovich convention, the $r=s=0$ is the Ito
convention. It is convenient to introduce a multi-index $\mu=(a=1,2,3; j=1,\ldots N)$ and rewrite
the Eq.~(\ref{eq:stoch1}) in the following form
\beq
i\dot{q}^{\mu}=-F^{\mu}({\bf q})+\sigma^{\nu}_{\mu}({\bf q}) \tilde{\phi}^{\mu},
\eeq
where $q^{\mu}=(n^{a})_{j}$, $F^{\mu}({\bf q})$ is a noise-free (deterministic part) of the
equations (which collects the terms containing the magnetic field $h^{a}$ and the source term ${\cal
J}(t)$, and $\sigma^{\mu}_{\nu}({\bf q})=(R^{a}_{\alpha})^{-1}_{k}O_{kj}$ is a matrix 
composed of MC forms $(R^{a}_{\alpha})_{j}$, and the transformation matrix $O_{kj}$, which puts the
noise $\phi_{j}^{a}$ into the form of the Gaussian normalized white noise, 
$\langle\tilde{\phi}^{\mu}(t)\tilde{\phi}^{\nu}(t')\rangle =\delta^{\mu\nu}\delta(t-t')$. 
This can
be always done as soon as the matrix $\Omega^{a,b}_{i,j}$ is non-degenerate. For our lattice spin
system the components of the matrix $O_{jk}$ are therefore made of the Fourier coefficients
normalized by the Fourier frequencies. The generalized vielbeins $\sigma^{\mu}_{\nu}$ normalized as
$\sigma^{\mu}_{\rho}\sigma_{\mu\nu}=\delta_{\rho\nu}$ define the metric tensor
$g_{\mu\nu}=\sigma_{\mu\rho}\sigma_{\nu\rho}$, so that $g^{\mu\nu}g_{\nu\rho}=\delta_{\nu\rho}$. The
effective continuous-in-time Lagrangian then takes the following form 
\beq 
{\cal L}_{r,s}[q,\dot{q}]
&=&\frac{1}{2}g_{\mu\nu}(\dot{q}^{\mu}+F^{\mu}(q))(\dot{q}^{\nu}+F^{\nu}(q))\nonumber\\ &+& s
\sigma_{\gamma\rho}\partial_{\nu}\sigma^{\nu}_{\rho}(\dot{q}^{\gamma}+F^{\gamma}(q))-r\partial_{\mu}F^{\mu}(q)\nonumber\\
&+&\frac{1}{2}s^{2}\partial_{\mu}\sigma^{\nu}_{\rho}\partial_{\nu}\sigma^{\mu}_{\rho}.
\label{eff-act-stoh}
\eeq 
This general form of the Lagrangian can be simplified for particular
physical systems. The integration measure in the path integral reads $D(q/\sqrt{\det{g_{\mu\nu}}})$.
This representation is invariant with respect to the change of the particular representation and/or
parametrization.  The partition function or the time-evolution operator for the lattice spin system
is than the expectation value of the operator $O_{N}=\prod_{j}U(q_{j})$, $W[{\cal J}]=\int
D(q/\sqrt{\det{g_{\mu\nu}}})O_N \exp(-\int dt {\cal L}_{r,s}[q,\dot{q}])$.

\subsection{Super-symmetry and connection to topological field theory}
We come back to the general discussion of certain global properties of our approach. We would like
to keep all stochastic trajectories without making any approximations. To do this we introduce an
identity for every lattice point (a customary trick in field theory~\cite{ZJ})
\beq
1=\int D[U] \det(M)\delta(J_{0}-\Phi),
\label{one}
\eeq
where we introduced the matrix notations for $\Phi\equiv \Phi^{a}S_{a}$, the current
$J_{0}=i\dot{U}U^{-1}$, and the Jacobian $M_{ab}=\delta J_{0}/\delta U_{ab}$ of the transformation.
Here  $d[U]$ is the Haar invariant integration measure on $G$, satisfying $\int d[U]f(U)=\int d[U]f(UV)$ for
any trace-like function~$f(U)$. Then $\delta(J_{0}-\Phi)=\int d\Lambda/(2\pi i)\exp[-Tr(\Lambda
(J_{0}-\Phi)]$ (where the integration is performed along the imaginary axis) defines dual fields
$\Lambda_{j}^{a}$, $\Lambda=\Lambda^{a}S_{a}$. Introducing the anti-commuting Faddev-Popov ghosts
$C^{a}, \bar{C}^{a}$ and using the Grassmann integration, one can write $\det M=\int d\bar C dC
\exp(\bar C^{a} M_{ab} C^{b})$ to obtain an effective generating functional after integrating out
the stochastic fields,
\beq
W[h]&=&\int d[U] d\Lambda d\bar{C}dC \exp(-\sum_{j}{\cal S}_{j})\prod_{j}\mbox{Tr}U_{j}\\
{\cal S}_{j}&=& Tr[\Lambda( J_{0}-h)]_{j} -\bar{C}_{j}D_{0}C_{j} +Tr(\Lambda_{j} \Omega_{jk}\Lambda_{k}).\nonumber
\label{susy-form}
\eeq
Here $D_{0}=\partial_{t}-J_{0}$ is the covariant derivative in time direction, $j$ is the lattice site
index,
and $C\equiv\{c^{a}\}$ represents a vector of Grassman variables. In this representation a generating
functional looks like an expectation value of the Polyakov-like string $\prod_{j}\mbox{Tr}U_{j}$ for a
theory described by the action $\sum_{j}S_{j}$. 
In principle, some of the variables in the effective action can be traced out explicitly. For
instance, the variables $\Lambda$ enter quadratically and can be integrated out; the corresponding effective
theory would look like the current-current interacting sigma-model coupled to fermions. The Grassmann
part can be shown to lead to a finite contribution, which can be computed explicitly. This eventually
produces the form (\ref{eff-act-stoh}) of the effective action since the Grassmann variables
propagator $\langle \bar{C}(t)C(t)\rangle=s$ where the parameter $0\leq s\leq 1$ is the discretization
introduced above. The group variables enter the action quadratically into  and can be partially
(because the expression for $U(n^{a})$ is not quadratic) integrated out. However at this point we
would like to keep all the variables as well as their duals to demonstrate an underlying {\it
super-symmetry}, which is difficult to see otherwise. 

The hidden BRST super-symmetry \cite{BRST} can be traced back to the stochastic nature of
differential equations (\ref{MC-eq}) or (\ref{eq:RiccatiEqsSU}) on the Lie group manifold. In this
sense it is a general feature of stochastic equations first noticed in \cite{PS},\cite{FT} (see also
\cite{ZJ}). The super-symmetry transformation $\delta$ on variables in (\ref{susy-form}) is induced
by
\beq
\delta\Lambda &=& 0, \qquad \delta\bar{C}=\epsilon\Lambda\\
\delta U &= & \epsilon UC,\qquad \delta C=-\epsilon C^{2},
\eeq
and as a consequence, $\delta J_{0}=\epsilon D_{0} C$ and $\delta(D_{0}C)=0$. In components it reads
$\delta c^{a}=-\frac{1}{2}f^{a}_{bc}\epsilon c^{b}c^{c}$. The action $\sum_{j}{\cal S}_{j}$ is
therefore BRST exact, $\delta{\cal S}=0$.  In terms of the group parameters $n^{a}$ the BRST charge
can be represented by the action of the operator 
\beq
{\cal Q}=\sum_{j}(c^{a}\delta_{n^{a}}+\Lambda_{a}\delta_{\bar{c}^{a}})_{j},
\eeq
so that ${\cal Q}^{2}=0$. Action on any functional is defined by $\{{\cal Q},X_{j}\}={\cal Q}X_{j}$.
One can further show that
\beq
{\cal S}&=&\{{\cal Q},V\}\\
V&=&\sum_{j}\bar{c}^{a}_{j}J_{0,j}^{a}+\sum_{i,j}\bar{c}^{a}_{i}\Omega_{ij}^{ab}\Lambda^{b}_{j}\\\nonumber
&=&
\sum_{j}\bar{c}^{a}_{j}J_{0,j}^{a}+\sum_{ij}\Omega_{ij}^{ab}\{Q,\bar{c}^{a}_{i}\{Q,\bar{c}^{b}_{j}\}\}.\nonumber
\label{top-action}
\eeq
In this form the action is in the form of the {\it topological field theory} of Witten
\cite{Witten}. It is easy to see that it is ${\cal Q}$-exact. The action of the ${\cal Q}$ operator
on $U$ is non-zero, and therefore the lattice spin-system partition function is equivalent to
computing expectation value of product of non-exact operators in this theory. Recently, progress in
computing "non-topological" (non-BPS) observables in topological field theories has been made in the
framework of instantonic filed theory \cite{FLN}.  Concrete applications of these developments to
the spin systems shall be discussed elsewhere. We emphasize that this representation of the
interacting spin system is exact, since we kept all the non-differentiable path-integral trajectories
exactly. 

There are several possible implications of the SUSY/topological field theory structure. First, there
are Ward-Takahashi identities associated with this super-symmetry. They lead to the
fluctuation-dissipation theorem and work relations \cite{work-relation}. Second, these identities
allow to establish  renormalization group prescriptions for the action. In particular we believe
that one can establish the RG flow equations using the generalized tetrads $\sigma^{\mu}_{\nu}$ in
the form of the geometric flow developed in  \cite{Friedan}. Third, there is a phenomenon of
dimensional reduction \cite{PS-prl}, coming from two additional Grassmann coordinates in the
effective action. Application of these phenomena to spin systems could lead to an interesting
relationships between spin systems defined in different dimensions. Fourth, a SUSY breaking pattern
is rather restrictive. There are in general three scenarios: unbroken SUSY, breaking on the mean field
level, and dynamical breaking. We believe that one may classify spin systems according to
these scenarios. Fifth, the formulation in terms of the topological field theory could shed a new
light on the origin of the topological order in certain spin systems. Moreover,
one of the central concept of the topological theory, the Witten index, seems to have direct
relevance to the sign problem \cite{Ovch}.

\section{Explicit demonstration of the method. Comparison with the Bethe ansatz and beyond}
We shall illustrate the technique discussed in the previous sections by considering a problem of a
single atom interacting with a one-dimensional photonic waveguide, treated within the Rotating Wave
Approximation (RWA). This problem has been solved by the Bethe ansatz in
\cite{ref:YudsonPRA}, see also \cite{ref:BeforeYudson,ref:ShiSunFan2,ref:ShiSunFan3}. 

\subsection{The model and the observables}
We assume that the photons in the
waveguide feature a linear dispersion with two (left,right) branches, described by annihilation
operators~$a_{R,L}$. For one scatterer only, it is convenient to work in the basis of even/odd
photons $a_{e,o} = (a_R\pm a_L)/\sqrt{2}$. Odd photons~$a_o$ do not interact with the atom,
which is why they evolve trivially as free photons. 
The interacting part of the evolution is hidden in the even component~$a_e$. 
From now on we focus purely on this nontrivial part and drop the index~$e$ ($a_e \equiv a$). 
We consider the interaction between a collection of $M$~atoms, located within a
region much smaller than the wavelength of light, and the even component of the photonic field. 
The atoms may be modeled as a single spin~$\op{S}$ of magnitude $j=M/2$.
The RWA Hamiltonian of the system then reads~\cite{ref:BeforeYudson,ref:YudsonPRA}
\begin{equation}
    \op{H} = \sum_k k \co ak \ao ak + \Delta(t) \op{S}^z 
    +  g(t) \sum_k  \left( \op{S}^+ \ao ak + \op{S}^- \co ak \right)
    \label{eq:GeneralHamiltonian}
\end{equation}
We have assumed $k$-independent, but potentially time-dependent light-matter coupling~$g(t)$, as
well as time dependent atomic detuning~$\Delta(t)$ (to save space, we in the following drop the
explicit time dependence from them).

Many interesting physical properties of this mode may
be expressed in terms of specific matrix elements of the evolution operator.
For the sake of calculations, we add the auxiliary photonic sources~$J^{(*)}_k(t)$, 
which serve the purpose of generating photons in the in- and out-states.
They are set to zero at the end of  the calculations. 
A generic matrix element of the evolution operator in presence of the photonic sources
reads
\begin{equation}
    \mathcal{U}_{oi}:=\braopket{\mathrm{out}}{\mathcal{T}\, e^{-i\int_{t_1}^{t_2} dt H +
    \sum_k \left( J^*_k(t) \ao ak +  J_k(t) \co ak \right).
    }}{\mathrm{in}}.
    \label{eq:GenFunZDef}
\end{equation}
We require that both $\ket{\mathrm{out}}$ and $\ket{\mathrm{in}}$ are tensor products of vacuum state
within the photonic subspace and a certain spin state. 
For instance, the amplitude of the decay of a fully excited state ($s^z=M/2$) into 
$2j$~photons can be related to Eq.~\eqref{eq:GenFunZDef} as
\begin{multline}
    T_{k_1,\dots,k_{2j}}:=\braopket{k_1,\dots,k_{2j};-j}{\mathcal{U}(\infty,0)}{0;j} =\\
    \frac{\partial^{2j}}{\partial J^*_{k_1}\dots\partial J^{*}_{k_{2j}}}
    \braopket{-j}{\mathcal{T}\, e^{-i\int_{0}^{\infty} dt H}}{j}.
    \label{eq:DecayAmplitudeDef}
\end{multline}
Also the problem of finding the scattering matrix of $n$~photons scattering on the atoms can be 
reduced to Eq.~\eqref{eq:GenFunZDef} using the LSZ formalism~\cite{ref:LSZ1,ref:ShiSunFan3}. This formalism expresses the
$T$-matrices of the scattering 
\begin{multline}
    iT_{k_i\rightarrow p_i} =
    (2\pi)^{n} \times \\
    G^{(2n)}(\{k_i\},\{p_i\})\prod_{s=1}^{n}\left(\mathcal{G}_{0}^{-1}(k_{s})\mathcal{G}_{0}^{-1}(p_{s})\right)|_{\mathrm{on-shell}}
    \label{eq:LSZFormula1}
\end{multline}
via the $n$-particle Green function, calculated as
\begin{equation}
    G^{(2n)}=\frac{(-1)^{n}\delta^{2n}\log Z[J,J^{*}]}{\delta
    J_{1}\ldots\delta J_{n}\delta J^{*}_{1}\ldots J^{*}_{n}},
    \label{eq:LSZFormula2}
\end{equation}
where~$Z[J,J^*]$ is the generating functional of the Green functions~\cite{ref:Negele} defined by
the equation
\begin{equation}
    Z[J,J^*] = 
    \braopket{-j}{\mathcal{T}\, e^{-i\int_{-\infty}^{\infty} dt H
    +\sum_k \left( J^*_k(t) \ao ak +  J_k(t) \co ak \right).
    }}{-j}.
    \label{eq:DefOfZ}
\end{equation}
This obviously forms a special case of Eq.~\eqref{eq:GenFunZDef}.

\subsection{Implementing disentangling transform and the stochastic formulation}
We may treat the amplitude~\eqref{eq:GenFunZDef} within the path-integral approach.
We rewrite the amplitude as a coherent state path integral over the 
photonic variables, while we retain the time-ordering for the spin variables~\cite{ref:Negele}
\begin{multline}
    \mathcal{U}_{oi} = \int \mathcal{D}[a,a^*] 
    \bra{\mathrm{out}} 
    \mathcal{T}\,
    e^{i\int dt 
    \sum_k a^*_k(i\partial_t-k)a_k
    }
    \times\\
    e^{i \int dt \left\{ \sum_k\left[ - g \left( a^*_k \op{S}^- +a_k \op{S}^+ \right) 
    + J^*_k a_k + J_k a_k^*
    \right]- \Delta \op{S}^z \right\}}
    \ket{\mathrm{in}}.
\end{multline}
The integral over the bosonic degrees of freedom may be performed explicitly and gives rise to
a spin model with an effective spin-spin interaction. Thanks to the linearity of the photonic
spectrum, this spin-spin interaction happens to be local in time, which simplifies the calculations
enormously. After some algebraic manipulations we arrive at the expression 
\begin{multline}
    \mathcal{U}_{oi} = \mathcal{U}^0_{oi}
    \bra{\mathrm{out}}
    \mathcal{T}\,
    e^{\int dt 
    -\frac{|g|^2}{2} {\op{\boldsymbol{S}}}^2 
    }\times
    \\
    e^{\int dt \left\{
    \frac{|g|^2}{2} {S^{z}}^2
    - \left[ i\Delta + \frac{|g|^2}{2} \right] S^z + \mathcal{J} S^+ +  \mathcal{J}^* S^-\right\}}
    \ket{\mathrm{in}},
    \label{eq:ZbeforeHS}
\end{multline}
with
\begin{equation}
    \mathcal{J}^{\mp}(t) := g(t)\sum_k \int_{0}^{\infty} d\tau  J_k^{(*)}(t\mp\tau)
    e^{-i k \tau}.
    \label{eq:defEffectiveSources}
\end{equation}
The symbol~$\mathcal{U}^0_{oi}$ denotes the free-photonic-evolution part of the amplitude, which is
irrelevant for the problems under consideration and consequently can be ignored.
As explained in the first part of this paper, the natural way of decoupling the spin-spin interaction
is via the Hubbard-Stratonovich transform (local in time)~\cite{ref:HSTransform}. 
The decoupling converts the exponent in Eq.~\eqref{eq:ZbeforeHS} into a linear form, disentangleable
using Eq.~\eqref{eq:DisentanglementSU}, 
at the expense of introducing a real-valued white-noise field~$\Phi(t)$:
\begin{multline}
    \mathcal{U}_{oi} = \mathcal{U}^0_{oi}
 \Big\langle   \bra{\mathrm{out}}
    \mathcal{T}\,
    e^{\int dt  -\frac{|g|^2}{2} \boldsymbol{S}^2 }\times
    \\
    e^{\int dt \left\{ 
    \left[ g \Phi - i\Delta - \frac{|g|^2}{2} \right] S^z + \mathcal{J}^- S^+ +  \mathcal{J}^+ S^-\right\}}
    \ket{\mathrm{in}}
    \Big\rangle_\Phi,
    \label{eq:ZafterHS}
\end{multline}
where we define the white-noise averaging~$\langle \dots \rangle_\Phi$ to be 
\footnote{In terms of the standard stochastic calculus 
the white noise $\Phi$~is related to the Wiener process~$W$ as $W(t)-W(s) = \int_s^t d\tau
\Phi(\tau)$~\cite{ref:Gardiner}.}
\begin{equation}
 \langle \mathcal{A}[\Phi] \rangle_\Phi = 
 \frac{
 \int\mathcal{D}[\Phi]
 e^{-\frac{1}{2}\int dt \Phi^2(t)} \mathcal{A}[\Phi]}{
\int\mathcal{D}[\Phi]
e^{-\frac{1}{2}\int dt \Phi^2(t)}}.
    \label{}
\end{equation}

The evolution operator in Eq.~\eqref{eq:ZafterHS}  can be now 
readily disentangled using the technology explained in the first part of this paper.
The evolution of the new variables~$\xi_{\pm,z}$ is governed by 
a specific form of Eqs.~\eqref{eq:RiccatiEqsSU}, which we converted into the Ito
form~\cite{ref:Gardiner}
\begin{gather}
    \dot \xi_+ = \mathcal{J}^- - i\Delta \xi_+ -\mathcal{J}^+ \xi_+^2 +  |g| \xi_+
    \Phi,
    \label{eq:DynSymQOStochRiccatiEqs}
    \\
    \nonumber   \dot \xi_z = -[i\Delta+{|g|^2/2}] - 2\mathcal{J}^+ \xi_+ + |g|  \Phi,
    \quad
    \dot \xi_- = \mathcal{J}^+ e^{\xi_z}.
\end{gather}
These equations  form a set of non-linear stochastic differential equations (SDE) in the Ito
form~\cite{ref:Gardiner}.\footnote{The equation Eqs.~\eqref{eq:RiccatiEqsSU} are written in the Stratonovich form. We
had to convert the equations into the Ito form, which is easier to work with.}
Because of their nonlinearity, it is not possible to write down their explicit solution.
We may, however, express the quantities of interest in terms of~$\xi_{\pm,z}$ and write down
the underlying SDE for these, utilizing the Ito lemma~\cite{ref:Gardiner}.\footnote{
The Ito lemma establishes the rule for the change of variables in a SDE. 
If a collection of stochastic functions~$x_i$ satisfy the SDEs
$d x_i = A_i dt + B_i dW$, then according to the Ito lema
a function~$f(x_i)$ satisfies the SDE
$d f = \left( \sum_i A_i \frac{\partial f}{\partial x_i} +
\sum_{i,j} \frac{1}{2}B_i B_j\frac{\partial^2 f}{\partial x_i \partial x_j} \right)dt + \sum_i B_i
\frac{\partial f}{\partial x_i} dW$.
}
Averaging these w.r.t. the noise~$\Phi$ we get rid of the noise term and arrive at ordinary
differential equations for the averages.
Inspecting the matrix elements of Eq.~\eqref{eq:DecayAmplitudeDef} 
and Eq.~\eqref{eq:DefOfZ} and comparing them to the disentangling transformation of
Eqs.~\eqref{eq:RiccatiEqsSU}, we conclude that these are straightforwardly related 
to the $\Phi$-averages of the functions~$\xi_-^{2j} e^{-j \xi_z}$, or $e^{-j\xi_z}$, respectively.
The Ito lemma applied to these,  however leads to unclosed equations, 
because extra powers of~$\xi_\pm$ pop up. 
This obstacle may be overcome by 
considering a more general problem, in particular a 
hierarchy of equations for the functions~$\xi_+^n e^{-j\xi_z}$, or $\xi_-^n e^{(j-n)\xi_z}$,
respectively. This 
includes the desired quantities as a special case. 
Averaging these equations as indicated above, we obtain a closed set of ordinary differential equations
for the following averages
\begin{equation}
    F_{n} := \av{\xi_+^{n} e^{-j \xi_z}}_\Phi, \quad
    R_{n} := \av{\xi_-^n e^{(j-n) \xi_z}}_\Phi,
    \label{eq:DynSymGenFunDef}
\end{equation}
with $0\le n\le 2j$.
The physically interesting quantities stem from $F_0$ and $R_{2j}$, respectively.
These auxiliary quantities satisfy by definition the initial conditions $F_0(t_1) = 1$,
$F_{n>0}(t_1) = 0$ (the same for $R_n$).
Based on Eqs.~\eqref{eq:DynSymQOStochRiccatiEqs} the quantities $F_n$ and $R_n$ satisfy a simple
system of linear ordinary differential equations (the noise has been already averaged out!)
\begin{multline}
    \dot{F_n} = \left[ i\Delta\left( j - n \right) + \frac{g^2}{2}
    \left( j\left( j + 1 \right) - 2j n +n (n-1)  \right) \right] F_n \\
    + n \mathcal{J}^-  F_{n-1}  + (2j-n) \mathcal{J}^+ F_{n+1},
    \label{eq:DynSymReducedHieararchy}
\end{multline}
\begin{equation}
    \dot R_n = [-i\Delta (j-n)  - \frac{g^2}{2}(2j-n)(n+1)] R_n +
    \mathcal{J}^+  n R_{n-1}.
    \label{eq:HierarchyDecay}
\end{equation}
These equations contain all information necessary to describe the two underlying physical problems.

\subsection{Scattering of $n$ photons: results}
According to Eq.~\eqref{eq:DefOfZ} and Eq.~\eqref{eq:DynSymGenFunDef} we can establish that all what
is needed to calculate the scattering matrix for any number of incoming photons is hidden in the
logarithm of the generating functional~$Z$ of Eq.~\eqref{eq:DefOfZ}, which up to terms that do not contribute to the scattering 
matrices reads
$\log Z[J,J^*]  = \log F_0(\infty)|_{t_1=-\infty}$.
It is now essential to realize that we can treat $\log Z$ perturbatively in the sources~$J^{(*)}$. 
According to the LSZ reduction formula~\eqref{eq:LSZFormula1}, the scattering of $n$-photons is 
fully and exactly described by the $2n$-th~term of the Taylor expansion of~$\log Z$ w.r.t the
sources~$J^{(*)}$. 
We introduce a control expansion parameter~$\lambda$, such that
$J^{(*)}=\mathcal{O}(\sqrt{\lambda})$, and calculate the expansion of~$Z[J,J*]=1+a_1+a_2+\dots$ with 
the coefficients  $a_n = \mathcal{O}(\lambda^n)$. The power series of $\log Z$ can be then written as 
$\log ( 1 + a_1 +  a_2 +  a_3 + \dots ) =
    a_1 + \left( a_2 - \frac{a_1^2}{2} \right)  + \left( a_3 - a_1 a_2 + \frac{a_1^3}{3} \right)
    + \dots$,
which we organized order by order. 

Solving the hierarchy~\eqref{eq:DynSymReducedHieararchy} up to 4th order in~$J^{(*)}$, we deduce
the $T$-matrices for 1- and 2-photon scattering~$T^{(1,2)}$,
\begin{equation}
    i T^{(1)}_{k\rightarrow p} = \delta(p-k) \frac{-2ijg^2}{k+ijg^2},
        \label{eq:TMatrix1PResult}
\end{equation}
and
\begin{multline}
    iT^{(2)}_{k_{1,2}\rightarrow p_{1,2}} = 
  \frac{  \delta(E-E')\cdot 16 g^4 i j E }{(E+ig^2(2j-1)) }\times
    \\
    \frac{ (E+2ijg^2)}{ [ (E+2ijg^2)^2 - \Delta^2 ][ (E+2ijg^2)^2 - \Delta'^2
    ] },
    \label{}
\end{multline}
After transforming the even/odd basis back to left/right basis, 
one may check that these results completely agree with those calculated using the Bethe ansatz 
in~\cite{ref:BeforeYudson,ref:YudsonPRA}.

\subsection{Decay of an initially excited state}
Let us investigate the problem of the decay of an initial fully excited state.
At $t_1=0$ all the emitters are supposed to be in the upper state, 
which corresponds to the highest weight spin state with~$s_z=j$.
The emitters consequently relax into the ground state and radiate outgoing photons. 
A conserved quantity ---  the number of excitations --- fixes
the number of outgoing photons to be~$2j$ ---  the number of atoms in the cluster. 
Equations \eqref{eq:DecayAmplitudeDef}, \eqref{eq:DynSymGenFunDef}, and \eqref{eq:HierarchyDecay} 
allow us to find an exact solution in quadratures for general sources~$J^{(*)}$.
For the time-independent coupling~$g$ and detuning~$\Delta$ the underlying integrals can be
performed analytically.
For instance, we may derive the spectral distribution of the outgoing photons, 
namely the single photon spectrum
\begin{equation}
    \mathcal{P}_{2j}(q) := \int \prod dk_m (\sum_m \delta(q-k_m)) |T_{k_1,\dots,k_{2j}}|^2.
    \label{eq:DecaySinglePhotonSpectrum}
\end{equation}
An explicit calculation gives a normalized result
\begin{equation}
    \mathcal{P}_{2j}(q) =   \sum_{m=0}^{2j-1} \frac{1}{2j\pi} \frac{g^2(j(j+1)-(j-m)^2)}{g^4 (j(j+1)-(j-m)^2)^2 +
    (q-\Delta)^2},
    \label{}
\end{equation}
where each of the terms is a normalized Lorentz curve
with the peak at $q=\Delta$ and the width~$j(j+1)-(j-m)^2$. 

\subsection{Time-dependent parameters}
Time-dependent parameters $\Delta(t)$ and $g(t)$ do not pose any fundamental obstacle for our
method, it is just not possible to obtain the results analytically. 
The time-dependent regime is, however, inaccessible to the Bethe ansatz solution, which suggest
superiority of the technique proposed here. On the other hand, 
a periodic driving of the coupling constant is of direct physical interest, for instance within 
the cavity-QED setups \cite{driven,ref:TimeDependentCouplingAndResonance}.

\begin{figure}
\begin{center}
    \includegraphics[width=0.45\textwidth]{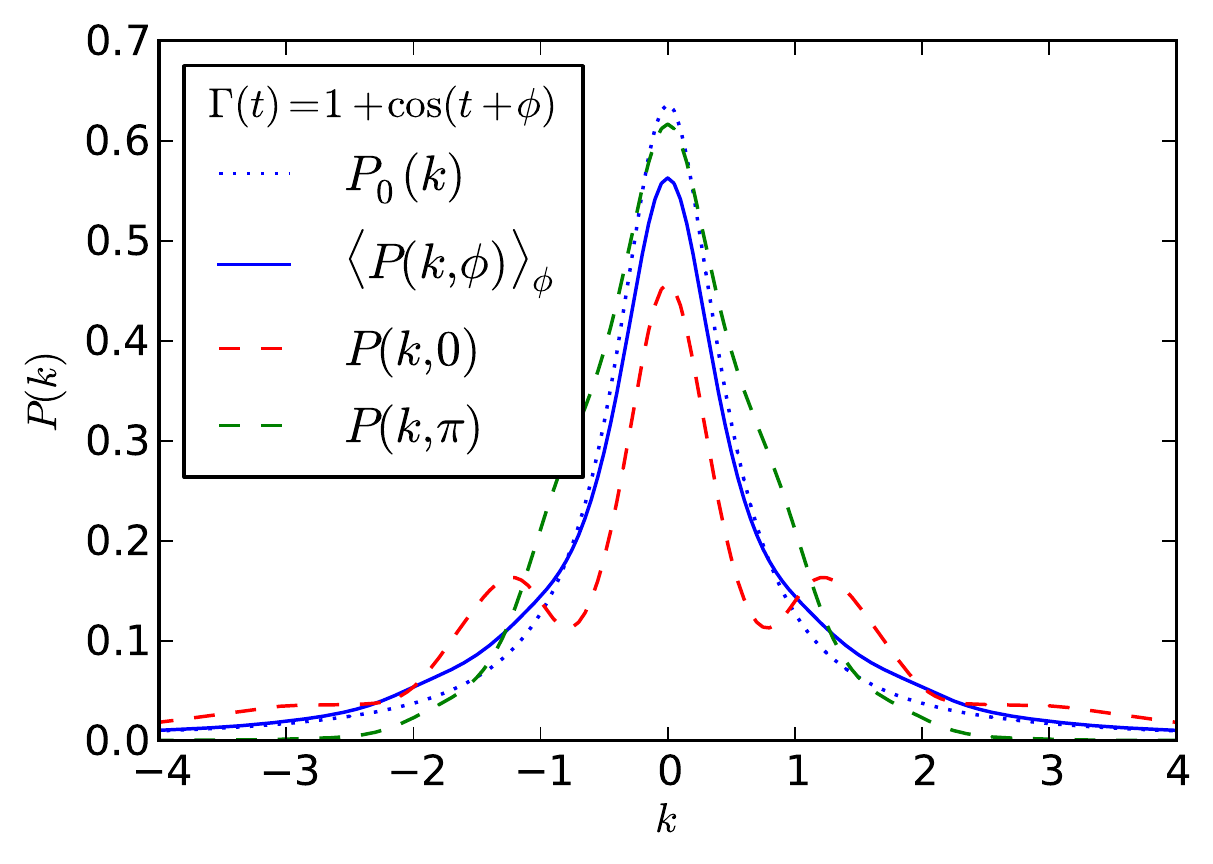}
    \includegraphics[width=0.45\textwidth]{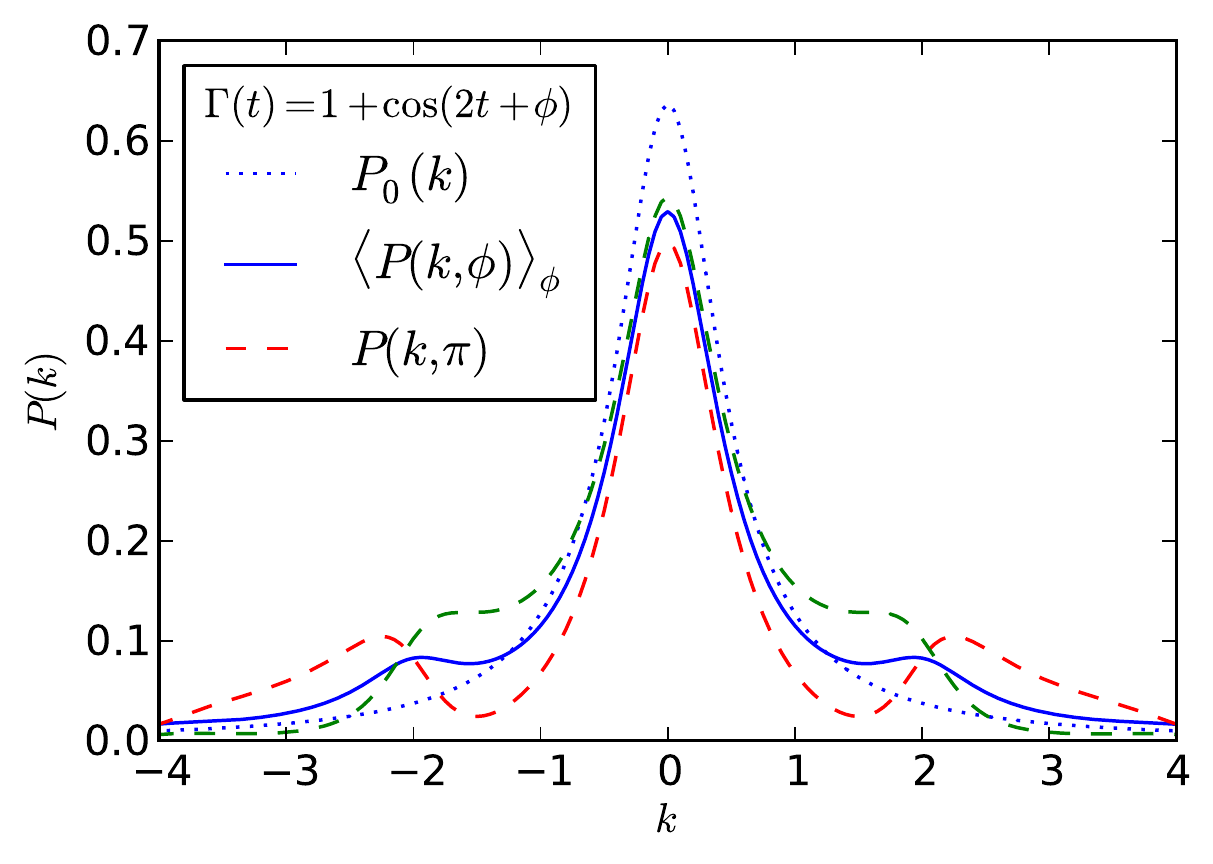}
    \includegraphics[width=0.45\textwidth]{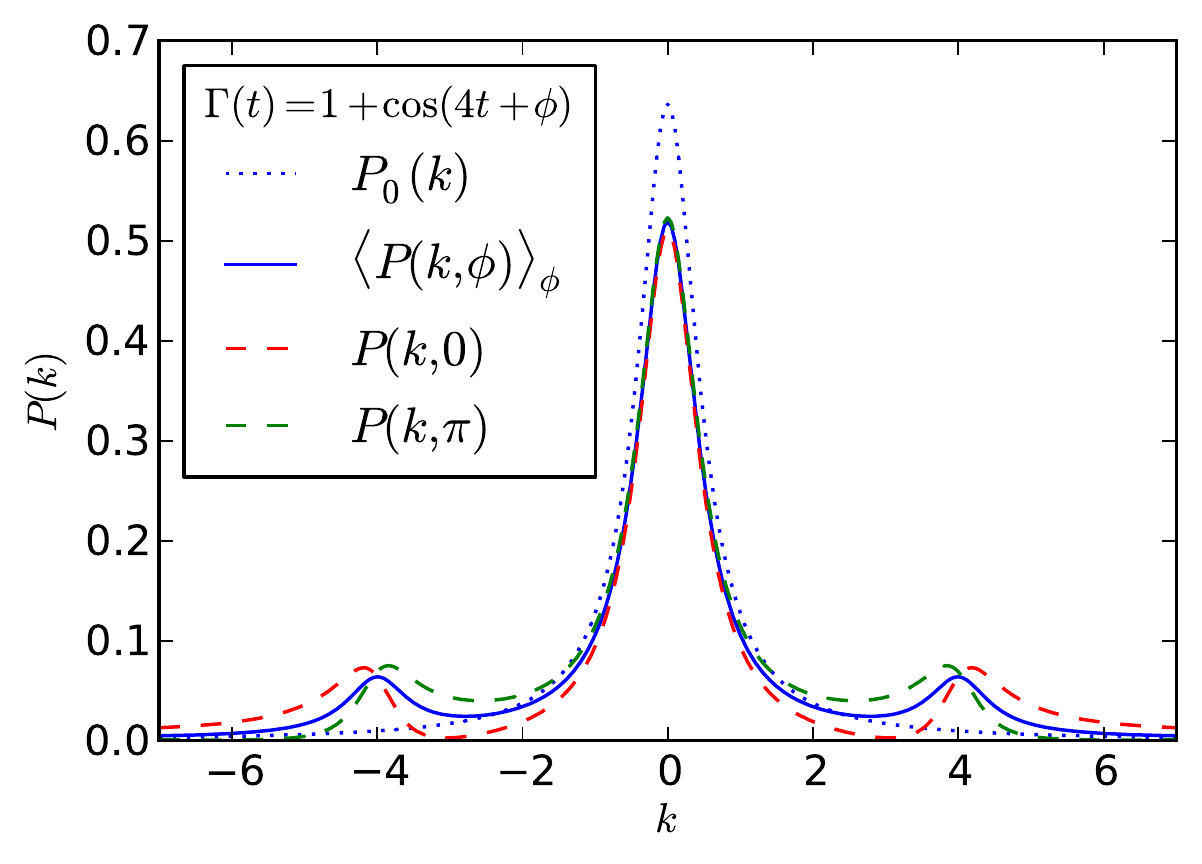}
\end{center}
\caption{A two-level system decays into one photon. 
We plot the probability distribution function~$P(k)$ of the photon's momentum~$k$, given
by Eq.~\eqref{eq:DecaySinglePhotonSpectrum} with $j=1/2$, i.e. $P(k) = \mathcal{P}_1(k)$.
The coupling strength $\Gamma$ is assumed to be time dependent, in particular oscillating, while
the effective level-splitting $\Delta$ is set to zero.
The probability distribution depends on the initial phase~$\phi$ of~$\Gamma(t)$. We plot~$P(k)$
averaged over $\phi$, its value for $\phi = 0, \pi$ and for comparison also the $\Gamma(t) = const.
=1$ probability distribution~$P_0(k)$.
Note that the oscillation $\Gamma(t) = 1+\cos(\Omega t+\phi)$ creates an additional peak at $k=\pm
\Omega$.   
}
\label{fig:twoLevelGoscDeltaConst}
\end{figure}

\begin{figure}
\begin{center}
    \includegraphics[width=0.45\textwidth]{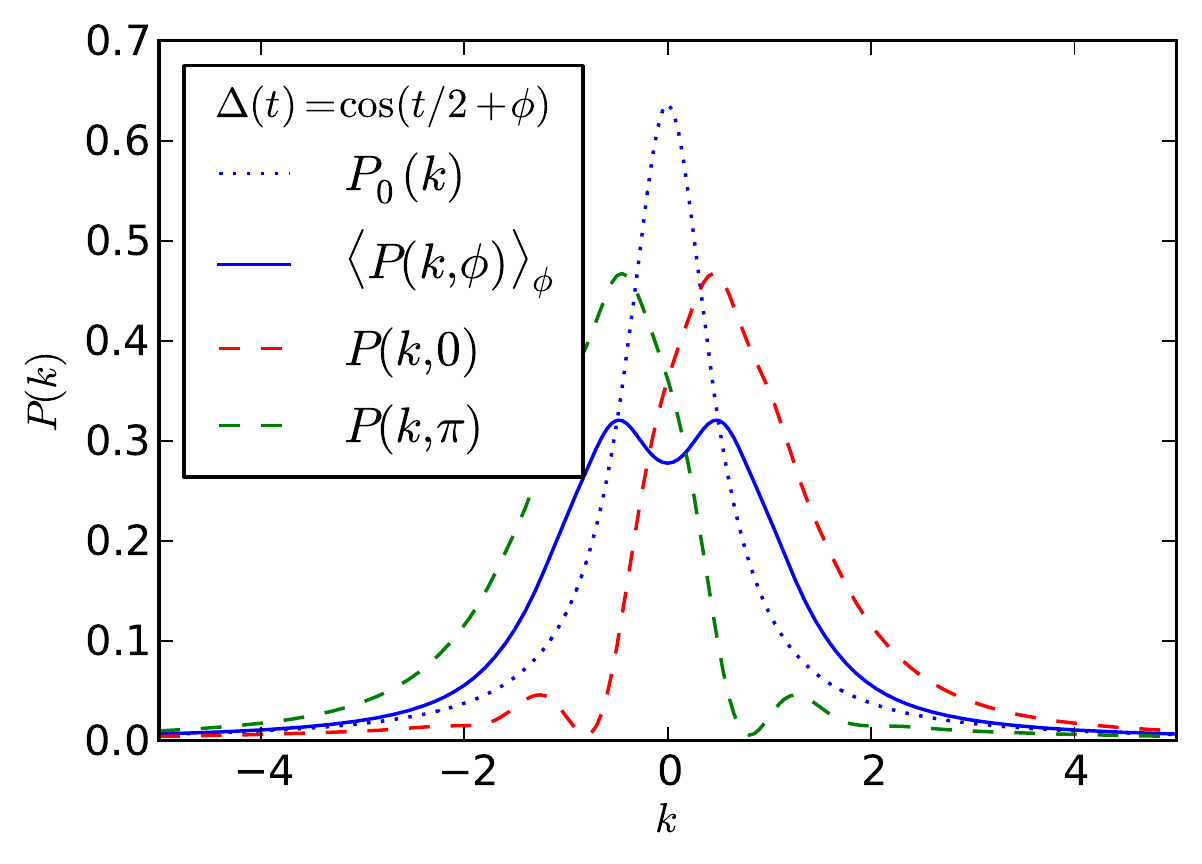}
    \includegraphics[width=0.45\textwidth]{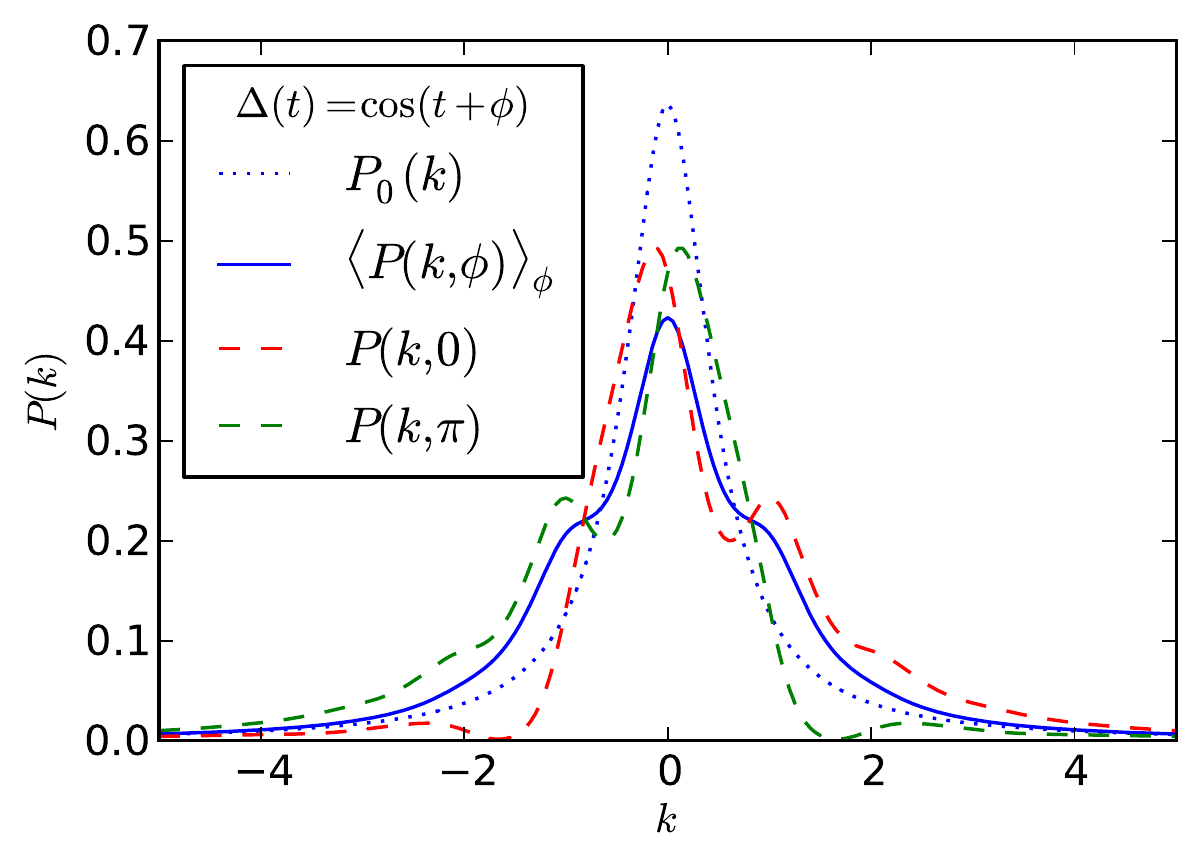}
    \includegraphics[width=0.45\textwidth]{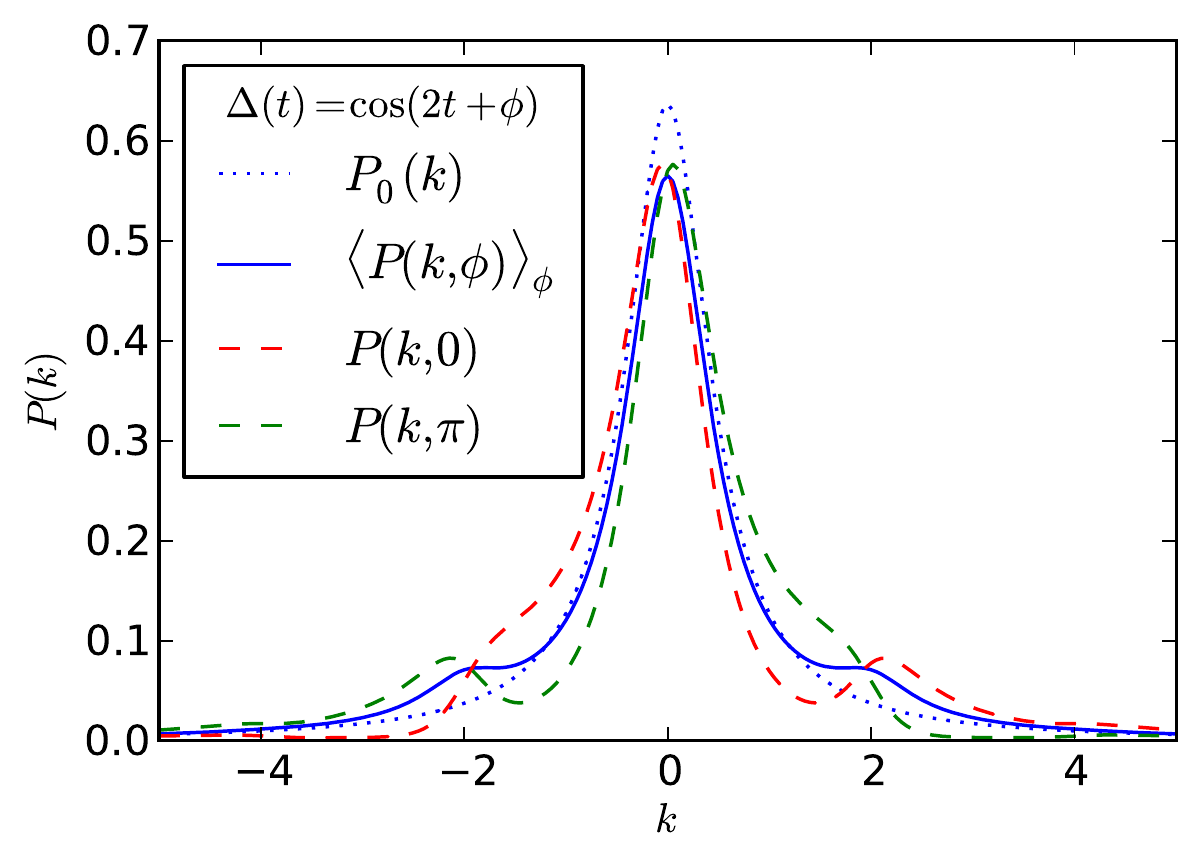}
\end{center}
\caption{A two-level system decays into one photon. 
We plot the probability distribution function~$P(k)$ of the photon's momentum~$k$, given
by Eq.~\eqref{eq:DecaySinglePhotonSpectrum} with $j=1/2$, i.e. $P(k) = \mathcal{P}_1(k)$.
The level-splitting $\Delta$ is assumed to be time dependent, in particular oscillating, while the
coupling constant~$g$ is assumed to be constant.
The probability distribution depends on the initial phase~$\phi$ of~$\Delta(t)$. We plot~$P(k)$
averaged over $\phi$, its value for $\phi = 0, \pi$ and for comparison also the $\Delta(t) = const.
=1$ probability distribution~$P_0(k)$.
 }
\label{fig:twoLevelDoscGConst}
\end{figure}

For illustration, we treat the hierarchy~\eqref{eq:HierarchyDecay} describing the decay of an
excited state numerically.
The solution stabilizes quickly with growing~$t_2$ and it is thus enough to retain only
a finite time $t_2$ to extract the limit~$t_2\rightarrow\infty$.
We explicitly demonstrate the effect of time-dependent parameters in the
case of a two-level system ($j=1/2$). 
We let the coupling constant~$\Gamma(t) = |g(t)|^2$ oscillate as $\Gamma(t) = \Gamma [ 1+ a
\cos(\Omega t + \phi)]$ and the atomic level splitting~$\Delta(t)$ to oscillate as~$\Delta(t) =
\Delta_0 \cos(\omega t + \varphi)$. Note that the amplitude depends on the initial phases $\phi$,
$\varphi$. As in experiment they may or may not be fixed, we present also a probablity distribution
averaged over possible initial phases.
The results of our numerical calculation are presented in
Figs.~\ref{fig:twoLevelGoscDeltaConst} and~\ref{fig:twoLevelDoscGConst}.
The oscillating coupling constant leads to an emergence of satellite peaks centered
at~$\pm\Omega$, whose strength decreases with a decreasing modulation depth~$a$. When $\Omega
\lesssim 1$ these satellite peaks are masked due to averaging over the initial phases, but they
survive the averaging when $\Omega$~is  larger.

\section{Discussion and Conclusion}

In this paper we discussed a path integral formulation for the lattice spin systems using the disentanglement
procedure of the time-ordered exponential. This approach avoids using coherent states 
and is therefore free of their drawbacks.  We invoked the stochastic formulation of the
disentanglement procedure which effectively keeps all the non-differentiable paths, and thus is in
principle exact. This allowed us to reproduce known results for a Bethe ansatz solvable model.
Moreover we used the method to describe time-dependent parameters where the Bethe ansatz
fails to provide a solution. From the fundamental point of view,
the stochastic approach to the disentanglement equations
reveals a hidden super-symmetric structure of the interacting spin system, and as a consequence
leads to the effective action formulation in terms of the topological field theory. We believe that
these new formulations could shed  light on hidden topological and dynamical aspects of interacting
lattice spin systems. 

We comment here on possible further developments of our representation. The Riccati equation which
plays a central role in the procedure has dynamical $sl(2)$ symmetry which could possibly help to
select certain orbits in the path integral. For systems defined on higher rank algebras,
such as~$su(n)$, this equation becomes a matrix Riccati equation.  The effective action in the
stochastic formulation can be studied using both non-perturbative and perturbative techniques. Many
of these approaches have been developed in the context of Onsager-Machlup theory for non-equilibrium
stochastic processes, see e.g. \cite{work-relation},\cite{ZJ}. On the other hand, our very
preliminary experience on numerical implementation of the method for generic spin system shows that
large deviations~\cite{ref:LargeDeviations1} will be an important obstacle to the numerical sampling of stochastic paths. 

One more interesting question is how the integrability of certain lattice spin model (like e.g. XYZ spin chains) can be uncovered via our path-integral formulation. We envision interesting developments on this route.  


\section{Acknowledgement}
We would like to thank Eugene Demler, Victor Galitski, Mikhail Kiselev, Mikhail Pletyukhov for many
useful discussions and suggestions. The work was supported by the Swiss NSF.

\section{Appendix}
In this appendix we collect some relevant information about several parametrizations of the $SU(2)$
group and related Riemann and differential geometry notions mentioned in the main text. 

\subsection{Different parametrizations of $SU(2)$}
Here we overview various parametrizations of the group $SU(2)$ and present
an explicit form of differential-geometric structures used in the main body of the paper. 
 
The {\bf Caley-Klein} parametrization
\beq
U(\alpha,\beta)=\left(
    \begin{array}{cc}
      \alpha & \beta \\
      -\bar{\beta} & \bar{\alpha} \\
    \end{array}
  \right),
\eeq
where $|\alpha|^{2}+|\beta|^{2}=1$. This allows one to introduce
new parameters $x_{1,\dots,4}$,
$\alpha=x_{1}+ix_{2}$ and $\beta=x_{3}+ix_{4}$, which define an
embedding of $S^{3}\simeq SU(2)$ to $R^{4}$. The underlying equations read
\beq
\phi_{3}&=&i(\dot{\alpha}\alpha+\dot{\beta}\beta) \\
\phi_{+}&=& i(-\dot{\alpha}\beta+\dot{\beta}\alpha),
\eeq
and the complex conjugates of these two equations.
Inverting them (when the determinant of the transformation is nonzero) yields
\beq
i\dot{\alpha}&=&\alpha\phi_{3}-\phi_{+}\bar{\beta}\\
i\dot{\beta}&=&\phi_{3}\beta+\phi_{+}\bar{\alpha}
\eeq

The {\bf Euler parametrization} is defined as
\beq
U(\alpha,\beta,\gamma)=e^{i\alpha(t)S^{3}}e^{i\beta(t)S^{2}}e^{i\gamma(t)S^{3}}.
\eeq
Here $0\leq\alpha<2\pi$, $0\leq\beta\leq \pi$, and
$0\leq\gamma\leq4\pi$. The stochastic equations take the following
form \cite{Chalker}
\beq
\dot{\alpha}&=&-\phi_{1}(t)\cos\alpha\cot\beta-\phi_{2}(t)\sin\alpha\cot\beta+\phi_{3}(t)\nonumber\\
\dot{\beta}&=&-\phi_{1}(t)\sin\alpha+\phi_{2}(t)\cos\alpha\\
\dot{\gamma}&=&\phi_{1}(t)\frac{\cos\alpha}{\sin\beta}+\phi_{2}(t)\frac{\sin\alpha}{\sin\beta}\nonumber.
\eeq
A {\bf rotationally invariant} (covariant) parametrization is given by
\beq
U(t)=\exp(-i\psi {\bf n}\cdot {\bf S}),
\eeq
where ${\bf n}(\theta,\phi)$ denotes  a unit vector spanning a sphere.  Its
connection to the Euler parametrization is obtained easily by
employing the lowest nontrivial representation --- the Pauli matrix  parametrization. 
It may be written as
\beq
\phi &=& \frac{1}{2}(\pi+\alpha-\gamma),\nonumber\\
\tan\theta &=&\frac{\tan(\beta/2)}{\sin[\frac{1}{2}(\alpha+\gamma)]},\\
\cos\psi &=&
2\cos^{2}(\frac{\beta}{2})\cos^{2}(\frac{1}{2}(\alpha+\gamma))-1.
\eeq

In this representation we define a vector ${\bf A}$  with three
components $(A^{1},A^{2},A^{3})$ (don't mix them up with parameters
$A,B,C$ in the text for the other representation), by requiring that
\beq
U(t)=\exp(i\bm{\sigma}{\bf A}(t))
\eeq
is the disentangled version of the $T$-ordered exponential. 
Using the standard formulas for the Pauli matrices we obtain
\beq
U(t)=\cos|A|+i\frac{\sin|A|}{|A|}(\bm{\sigma}{\bf A})
\eeq
where $|A|$ is the length of the vector ${\bf A}$. Considering the
time derivative $\partial_{t}U(t)$ and multiplying it by 
$U^{-1}(t)$ we get
\beq
-iU^{-1}\partial_{t}U(t)&=&\frac{|A|-\cos|A|\sin|A|}{|A|^{2}}\partial_{t}|A|(\bm{\sigma}\cdot{\bf
A})\nonumber\\
&+&\frac{\cos|A|\sin|A|}{|A|}(\bm{\sigma}\cdot\partial_{t}{\bf
A})\\
&+&\frac{\sin^{2}|A|}{|A|^{2}}(\bm{\sigma}\cdot({\bf
A}\times\partial_{t}{\bf A}))\nonumber,
\eeq
which is to be then identified with the effective Hamiltonian
$\bm{\phi}(t)\cdot\bm{\sigma}$. We therefore get a vector Langevin
equation
\beq\label{covariant-rep}
\bm{\phi} &=& \frac{|A|-\cos|A|\sin|A|}{|A|^{2}}\partial_{t}|A|{\bf
A}\\
&+&\frac{\cos|A|\sin|A|}{|A|}\partial_{t}{\bf
A}+\frac{\sin^{2}|A|}{|A|^{2}}({\bf A}\times\partial_{t}{\bf A}).\nonumber
\eeq
Note also that $(\bm{\phi}\cdot{\bf A})=\frac{1}{2}\partial_{t}|A|^{2}$.
From this form it follows that the leading terms of a power expansion in terms of 
$|A|$ read $\bm{\phi}=\partial_{t}{\bf A}+{\bf A}\times\partial_{t}{\bf A}+\ldots$

The connection between different representations can be obtained by
comparing their matrix forms with the one of the Caley-Klein:
\beq
\alpha &=& \cos\frac{|{\bf A}|}{2}+i\frac{A^{z}}{|{\bf A}|}\sin|{\bf
A}|\\
\beta &=& i\frac{A_{x}-iA_{y}}{|{\bf A}}\sin|{\bf A}|,
\eeq
while for the Gauss parametrization, $\exp(-B/2)=\bar{\alpha}$.

\subsubsection{Various disentangling formulas}
Let $S_{\pm,0}$ be the generators of $su(2)$ or $su(1,1)$, so that $[S_{-},S_{+}]=2\sigma S_{0}$
and $[S_{0},S_{\pm}]=\pm S_{\pm}$, where $\sigma=1$ for the $su(1,1)$ and $\sigma=-1$ for the
$su(2)$. The Casimir invariant is therefore
$C_{2}=S_{0}^{2}-\frac{\sigma}{2}(S_{+}S_{-}+S_{-}S_{+})$. The evolution operator can be written in
a variety of forms  distinguished by a different ordering of terms containing $S_{\pm,0}$.
These include the symmetric form,
\beq
U_{S}=\exp(a_{+}S_{+}+a_{0}S_{0}+a_{-}S_{-})
\eeq
the normal-ordered Cartan form
\beq
U_{N}=\exp(A_{+}S_{+})\exp(\log[A_{0}]S_{0})\exp(A_{-}S_{-})
\eeq
or the anti-normal-ordered Cartan form
\beq
U_{A}=\exp(B_{-}S_{-})\exp(\log[B_{0}]S_{0})\exp(B_{+}S_{+}).
\eeq
They are related as follows~\cite{ban}:
\beq
A_{\pm}&=&\frac{\frac{a_{\pm}}{D}\sinh D}{\cosh
D-\frac{a_{0}}{2D}\sinh D},\nonumber\\
A_{0} &=& \left(\cosh
D-\frac{a_{0}}{2D}\sinh D\right)^{2}\nonumber\\
B_{\pm} &=& \frac{\frac{a_{\pm}}{D}\sinh D}{\cosh
D+\frac{a_{0}}{2D}\sinh D},\nonumber\\
B_{0} &=& \left(\cosh
D+\frac{a_{0}}{2D}\sinh D\right)^{2}
\eeq
and
\beq
A_{0}=\frac{B_{0}}{(1-\sigma B_{+}B_{-}B_{0})^{2}},\quad
A_{\pm}=\frac{B_{\pm}B_{0}}{1-\sigma B_{+}B_{-}B_{0}}\nonumber\\
B_{0}=\frac{(A_{0}-\sigma A_{+}A_{-})^{2}}{A_{0}},\quad
B_{\pm}=\frac{A_{\pm}}{A_{0}-\sigma A_{+}A_{-}}
\eeq
where $D=\frac{1}{2}(a_{0}^{2}-4\sigma a_{+}a_{-})^{1/2}$.

The ordered product of arbitrary number of symmetric operators,
$U_{S}(n)=\exp(a_{+}(n)J_{+}+a_{0}(n)J_{0}+a_{-}(n)J_{-})$ labeled
by some discrete index $n=1,\ldots N$,
\beq
U_{S}(N)=\prod_{n=1}^{N}U_{S}(n)=U_{S}(N)U_{S}(N-1)\ldots U_{S}(1)
\label{discrete}
\eeq
can be represented in any of the forms above. In particular, for any $n$ the
normal ordered form 
\beq
U_{S}(n)=\exp(A_{+}(n)S_{+})\exp(A_{0}(n)S_{0})\exp(A_{-}(n)S_{-}),\nonumber
\eeq
can be obtained recursively using the quantities $A_{\pm,0}(n-1)$ defined at the previous
discretization step $n-1$ and the  quantities $a_{\pm,0}(N)$ defined at the step $n$. The explicit
formulas are stated in \cite{ban}. 


\subsection{Connection between group-theoretic and geometric approaches: example of $SU(2)$}
For the group $SU(2)$ we define $[t^{a},t^{b}]=i\epsilon^{abc}t^{c}$. A
basis for the algebra can be taken to be a set of Pauli matrices. We
have for the arbitrary element of $g\in G$,
$g^{-1}dg=\sum_{a}\omega_{a}t_{a}$, where $\omega_{a}$ is the
Maurer-Cartan form. These 1-forms satisfy
$d\omega_{a}+\frac{1}{2}f_{abc}\omega_{b}\wedge\omega_{c}=0$,
because of the identity $d(g^{-1}dg+g^{-1}dg\wedge g^{-1}dg=0$.
Parametrizing $SU(2)$ as
$\alpha=\cos(u_{1}/2)\exp(i(u_{2}+u_{3})/2)$,
$\beta=i\sin(u_{1}/2)\exp(i(u_{2}-u_{3})/2)$, the 1-form becomes
\beq
\frac{i}{2}\left(
  \begin{array}{cc}
    du_{3}+\cos u_{1}du_{1} & e^{-iu_{3}}(du_{1}+idu_{2}\sin u_{1} \\
    e^{iu_{3}}(du_{1}-idu_{2}\sin u_{1} & -du_{3}-\cos u_{1}du_{1} \\
  \end{array}
\right).
\eeq
This  means that
\beq
\omega_{1}&=&\cos u_{3}du_{1}+\sin u_{3}\sin u_{1} du_{2}\nonumber\\
\omega_{2}&=&\sin u_{3}du_{1}-\cos u_{3}\sin u_{1} du_{2}\nonumber\\
\omega_{3}&=&\cos u_{1}du_{2}+du_{3}.
\eeq
Using these equations, one can readily check the Maurer-Cartan
equation,
$d\omega_{a}+\frac{1}{2}\epsilon_{abc}\omega_{b}\wedge\omega_{c}$.

Geometrically speaking, $SU(2)$ is equivalent to $S^{3}$. The metric
induced on it by the embedding into $R^{4}=C^{2}$ can be found
easily,
\beq
ds^{2}=du_{1}^{2}+du_{2}^{2}+du_{3}^{2}+2\cos u_{1} du_{2}du_{3}.
\eeq
The determinant of the metric $\det(g_{\mu\nu})=sin^{2} u_{1}$ and
therefore the volume of the $S^{3}$ is normalized as
$\int_{SU(2)}\det^{1/2}(g_{\mu\nu})du_{1}du_{2}du_{3}=16\pi^{2}$,
the inverse of the metric yields
\beq
g^{\mu\nu}=\left(
  \begin{array}{ccc}
    1 & 0 & 0 \\
    0 & \csc^{2}u_{1} & -\cot u_{1}\csc u_{1} \\
    0 & - \cot u_{1}\csc u_{1} & \csc^{2} u_{1} \\
  \end{array}
\right),
\eeq
and nonzero Christoffel symbols read
\beq
\Gamma^{1}_{23}&=&\frac{1}{2}\sin u_{1},\qquad
\Gamma^{2}_{13}=\Gamma^{3}{12}=\frac{-1}{2\sin u_{1}},\nonumber\\
\Gamma^{3}_{13}&=&\Gamma^{2}{12}=\frac{1}{2}\cot u_{1}.
\eeq
The dreibein on $S^{3}$ is equal to $\omega$ within our normalization
(otherwise there may arise a proportionality factor),
\beq
e^{a}_{\mu}=(\omega_{a})_{\mu},\qquad
e^{a}=e^{a}_{\mu}dx^{\mu}=\omega_{a}.
\eeq
One can check that
\beq
e^{a}_{\mu}e^{b}_{\nu}\eta_{ab}=g_{\mu\nu}.
\eeq
The inverse dreibein is
\beq
E^{\mu}_{a}=\eta_{ab}g^{\mu\nu}e^{b}_{\mu}
\eeq
and is used to define left-invariant vector fields
\beq
l_{a}=E^{\mu}_{a}\frac{\partial}{\partial x^{\mu}},
\eeq
which in this case are
\beq
l_{1}&=&\cos u_{3}\frac{\partial}{\partial u_{1}}+\frac{\sin
u_{3}}{\sin u_{1}}\frac{\partial}{\partial u_{2}}-\sin u_{3}\cot
u_{1}\frac{\partial}{\partial u_{3}}\nonumber\\
l_{2}&=&\sin u_{3}\frac{\partial}{\partial u_{1}}-\frac{\cos
u_{3}}{\sin u_{1}}\frac{\partial}{\partial u_{2}}+\cos u_{3}\cot
u_{1}\frac{\partial}{\partial u_{3}}\nonumber\\
l^{3}&=&\frac{\partial}{\partial u_{3}}.
\eeq
One can see that $e^{a}(l_{b})=\delta^{a}_{b}$ and that these vector
fields form the $su(2)$ algebra themselves. The spin connection
satisfies $de^{a}+\omega^{a}_{b}\wedge e^{b}=0$ and is given
explicitly by equations
\beq
\omega^{a}_{b\mu}&=&-E^{\nu}_{b}(\partial_{\mu}e^{a}_{\nu}-\Gamma^{\lambda}_{\mu\nu}e^{a}_{\lambda}),\nonumber\\
\partial_{\mu}e^{a}_{\nu}&=&\Gamma^{\lambda}_{\mu\nu}e^{a}_{\lambda}-e^{b}_{\nu}\omega^{a}_{b\mu}\\
\omega^{a}_{b}&=&\epsilon^{a}_{bc}e^{c}.\nonumber
\eeq
The Laplace-Beltrami operator yields $\Delta_{LB}=l_{a}^{2}$.

All the expressions above are representation-dependent. In
particular, in the representation defined by the Cartan
decomposition we have
\beq
g_{G}&=&e^{\psi S_{-}}e^{\phi S_{z}}e^{\chi S_{+}}\\
&=&\left(
                                                     \begin{array}{cc}
                                                       1 & 0 \\
                                                       \psi & 1 \\
                                                     \end{array}
                                                   \right)\left(
                                                            \begin{array}{cc}
                                                              e^{\phi} & 0 \\
                                                              0 & e^{-\phi} \\
                                                            \end{array}
                                                          \right)\left(
                                                                   \begin{array}{cc}
                                                                     1 & \chi \\
                                                                     0 & 1 \\
                                                                   \end{array}
                                                                 \right)\nonumber,
\eeq
and the Maurer-Cartan currents are given by
\beq
g^{-1}dg &=& \left(
           \begin{array}{cc}
             -e^{2\phi}\chi d\psi +d\phi& -e^{2\phi}\chi^{2}d\psi+2\chi d\phi+d\chi \\
             e^{2\phi}d\psi & e^{2\phi}\chi d\psi-d\phi \\
           \end{array}
         \right)\nonumber,\\
         dg g^{-1} &=&\left(
           \begin{array}{cc}
             -e^{2\phi}\psi d\chi +d\phi &  e^{2\phi}d\chi\\
             -e^{2\phi}\psi^{2}d\chi+2\psi d\phi+d\psi & e^{2\phi}\psi d\chi -d\phi \\
           \end{array}
         \right),\nonumber
\eeq
while the left-invariant vector fields read
\beq
l_{+} &=&\frac{\partial}{\partial\chi},\quad
l_{3} = -2\chi\frac{\partial}{\partial\chi}+\frac{\partial}{\partial\phi},\nonumber,\\
l_{-} &=& e^{-2\phi}\frac{\partial}{\partial\psi}+\chi\frac{\partial}{\partial\phi}-\chi^{2}\frac{\partial}{\partial\chi}
\eeq
and the right-invariant fields yield
\beq
r_{+} &=&\frac{\partial}{\partial\psi},\quad
r_{3} = -2\psi\frac{\partial}{\partial\psi}+\frac{\partial}{\partial\phi},\nonumber,\\
r_{-} &=& e^{-2\phi}\frac{\partial}{\partial\chi}+\psi\frac{\partial}{\partial\phi}-\psi^{2}\frac{\partial}{\partial\psi}.
\eeq
The components of the vielbein can be read off these results straightforwardly. Validity
of the Maurer-Cartan equation can also be checked. Also note that
$(g^{-1}dg)^{a}r_{b}=\delta^{a}_{b}$, and $(dg g^{-1})^{a}l_{b}=\delta^{a}_{b}$.

\subsection{Discretized version of the disentangling equations in the stochastic formulation} 
If one wants to interpret the equations as stochastic differential equations with respect to the
Wiener process, one should look into the exact discretization outlined in Eqs.~(\ref{discrete}), and
explicitly in Ref.~\cite{ban}. Let us split~$\phi_{\pm,3}$ into a white-noise and deterministic
part, i.e. define $\phi_{\pm,3} = \sqrt{\varepsilon} \varphi_{\pm,3} + \varepsilon h_{\pm,3} =
dR_{\pm,3} + h_{\pm,3} dt$.  For convenience we denote the discretized versions of
$\xi_{+}(t),\xi_{-}(t), \xi_{z}(t)$ as $A_n, C_{n}, B_n$, respectively. The discretized 
equations then take the following form:
\beq
    A_{n+1} &=& A_{n} +  \sqrt{\varepsilon} \left( \varphi_{+,n} + A_n \varphi_{3,n} - A_n^2
    \varphi_{-,n} \right) \nonumber\\
    &+&
    \varepsilon \left( h_{+,n}  + A_n h_{3,n} - A_n^2  h_{-,n} \right) +\nonumber\\
    &+&
    \frac{\varepsilon}{2} ( \varphi_{+,n} \varphi_{3,n} + A_n \varphi_{3,n}^2 -2 A_n
    \varphi_{+,n} \varphi_{-,n} \nonumber\\
    &-& 3A_n^2 \varphi_{3,n} \varphi_{-,n}
     + 2 A_n^3 \varphi_{-,n}^2 )  +
    o(\varepsilon)\nonumber\\
    C_{n+1} &=& C_n + \sqrt{\varepsilon} e^{B_n} \varphi_{-,n} \\
    &+&
    \varepsilon e^{B_n} \left( h_{-,n} + \frac{1}{2} \varphi_{3,n} \varphi_{-,n} - A_n
    \varphi_{-,n}^2 \right) + o(\varepsilon)\nonumber\\
    e^{B_{n+1}} &=& e^{B_n} + \sqrt{\varepsilon} e^{B_n} ( \varphi_{3,n} - 2 A_n \varphi_{-,n} )\nonumber\\
    &+& \varepsilon e^{B_n} ( h_{3,n} -2 A_n h_{-,n} + \frac{1}{2} \varphi_{3,n}^2 -
    \varphi_{+,n} \varphi_{-,n} \nonumber\\
    &-& 3 A_n \varphi_{3,n} \varphi_{-,n} + 3A_n^2 \varphi_{-,n}^2  ) +
    o(\varepsilon)\nonumber.
\eeq
These equations form the basis for the construction of the corresponding~SDEs. The particular form
of the equations clearly depends on the particular properties of the noises.
For a general model with the quadratic coupling~$\frac{1}{2}\Omega_{ij}\left(\gamma_x S^x_i S^x_j
+ \gamma_y S^y_i S^y_j + \gamma_z S^z_i S^z_j\right)$ the noise action is $\frac12\Omega^{-1}_{ij}
(\gamma^{-1}_x \varphi^x_i
\varphi^x_j +
\gamma^{-1}_y \varphi^y_i\varphi^y_j  + \gamma^{-1}_z \varphi^z_i \varphi^z_j )$.
The noises are thus~$\varphi_{\pm} = \frac{1}{2} ( \varphi_x \mp i
 \varphi_y )$, $\varphi_3 = \varphi_z$.
If the noises~$dR_{\pm,3}$ are independent, we can rewrite the relations in the form of Ito SDE,
\beq
 dA &=& ( h_+ + A h_3 - h_- A^2 \nonumber\\
 &-& \frac{K}{4} A ( \gamma_x(1-A^2) +
    \gamma_{y}(1+A^2) - 2\gamma_z)) dt \nonumber\\
    &+&
    \frac12 (1-A^2) d R_x - \frac i2 (1+A^2)dR_y + A dR_3\nonumber\\
    \label{eq:ComplicatedEquations}
    d C &=& e^{B} ( h_- - \frac{K}{4} A ( \gamma_x - \gamma_y )
    ) dt \\
    &+& \frac 12 e^{B} (  dR_x + i  dR_y )     \nonumber\\
    de^B &=& e^B ( h_3 -2 A h_-\nonumber \\
    &+& \frac{K}{4} (2\gamma_z + \gamma_x(3A^2-1) -
    \gamma_y (3A^2+1)) ) dt \nonumber\\
    &+&
    e^B ( dR_3 -  A (dR_x + i dR_y )),\nonumber
\eeq
Here we assume that the noises~$dR$ can be written in terms of unit
Wiener processes~$dW$ via~$dR_{x,y,z} = \sqrt{\gamma_{x,y,z}} \sum_i T_{il} dW_{l;x,y,z}$. The
constant $K$ in Eqs.~\eqref{eq:ComplicatedEquations} is expressed via~$T_{ij}$ as~$K:=\sum_l T_{il}^2$.
The matrix~$T_{ij}$ can be straightforwardly related to the inverse of the
matrix~$O_{ij}$ defined above Eq.~\eqref{eff-act-stoh}.
These equations can be casted into a much simpler form by utilizing the Stratonovich interpretation:
\begin{gather}
    dA = \left( h_+ + A h_3 - h_- A^2 \right) dt\nonumber \\+
    \frac12 (1-A^2) d R_x - \frac i2 (1+A^2)dR_y + A dR_3\nonumber\\
    d C = e^{B} \left( h_-
    \right) dt + \frac 12 e^{B} \left(  dR_x + i  dR_y \right)     \\
    de^B = e^B \left( h_3 -2 A h_-  \right) dt +
    e^B \left( dR_3 -  A (dR_x + i dR_y \right)).\nonumber
\end{gather}

\begin{figure}[tb]
\begin{center}
    \includegraphics[width=0.48\textwidth]{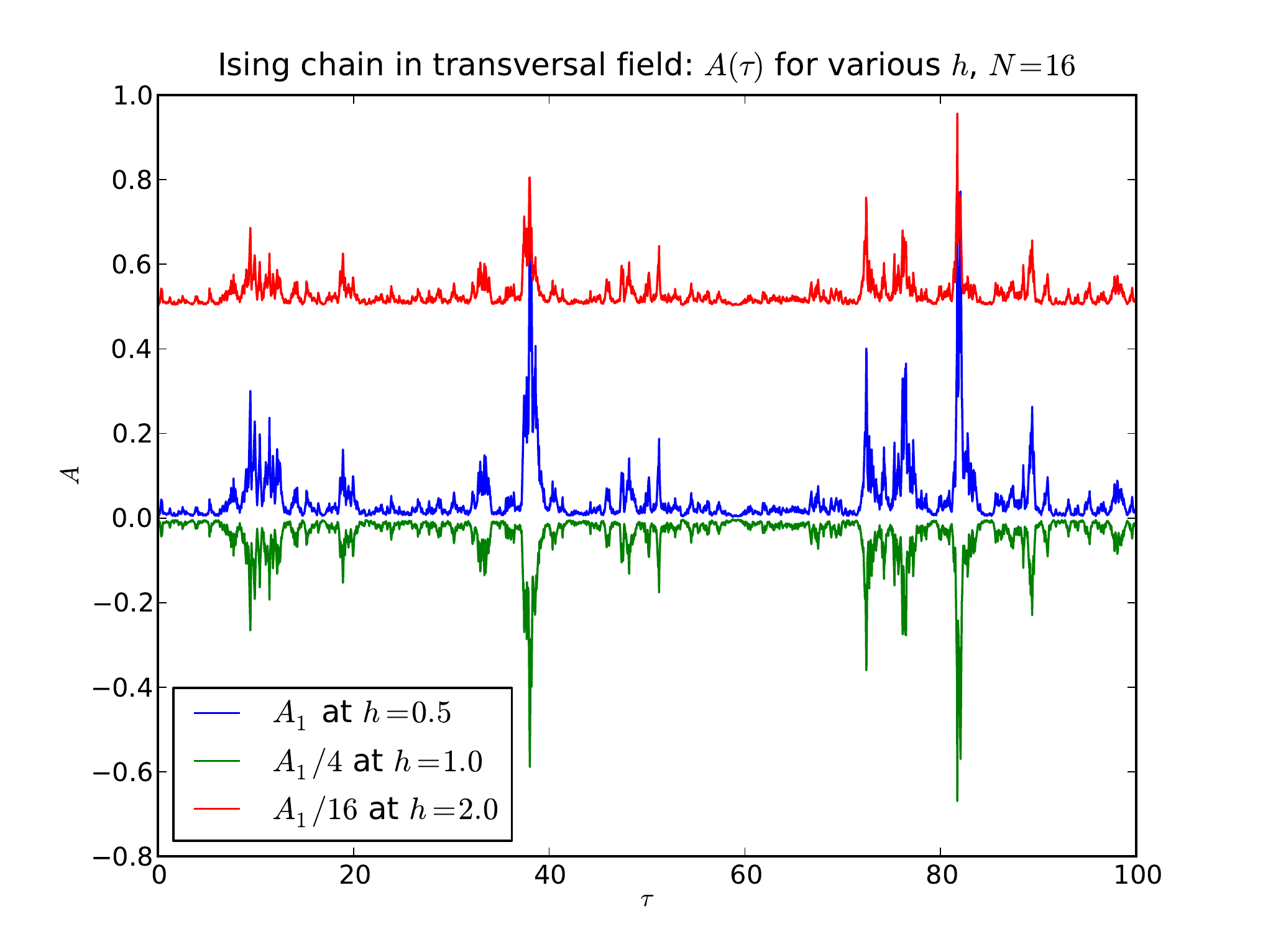}
    \includegraphics[width=0.48\textwidth]{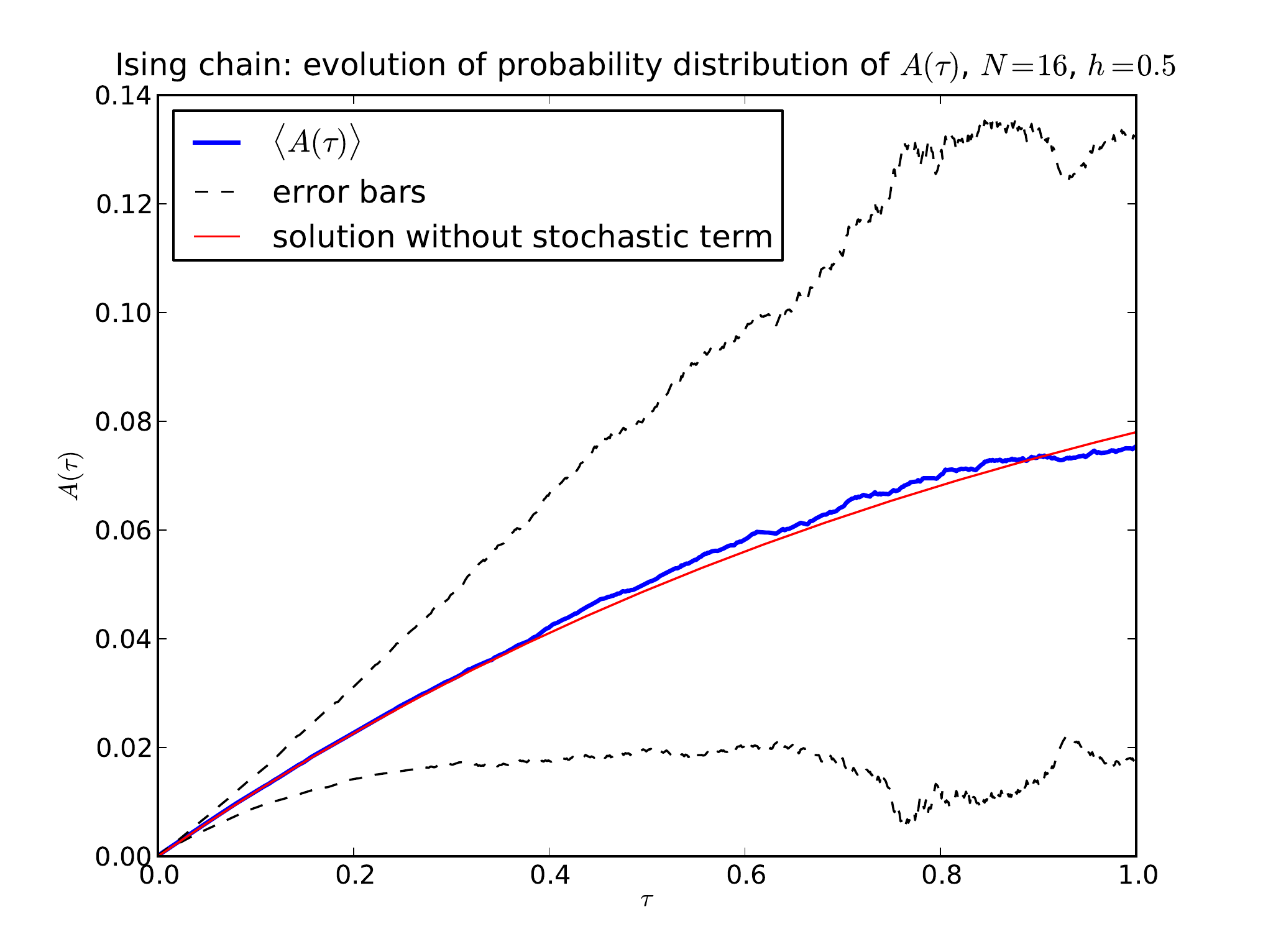}
\end{center}
\caption{ {\bf Upper panel:} Typical realizations of the stochastic process~$A_1(\tau)$,  in imaginary
time~$\tau$ for various magnetic fields~$h$. Values of~$A_1$ for different~$h$s 
are rescaled, mirrored and shifted to improve their visibility. 
The magnitude of~$A$ scales approximately as~$h^2$. 
Trajectories are all taken for the same noise realization, so that one can
compare the influence of a fluctuation in noise on~the solution~$A$ at different~$h$.  
{\bf Lower panel:}
Initial part of the evolution of~$A_1(\tau)$. The red line is a
solution of the underlying SDEs 
neglecting the contributions of the noise  term.  
The blue line is an average of solutions, averaged over
1000 realizations of the noise. 
Dashed lines represent error bounds given by~$\sqrt{\mathrm{var} A(\tau)}$. We can clearly see the
crossover from the ``deterministic'' region at small~$\tau$ into a purely stochastic stationary
regime at larger~$\tau$.
}
\label{fig:HistogramsAs1}
\end{figure}

\begin{figure}[tb]
\begin{center}
    \includegraphics[width=0.48\textwidth]{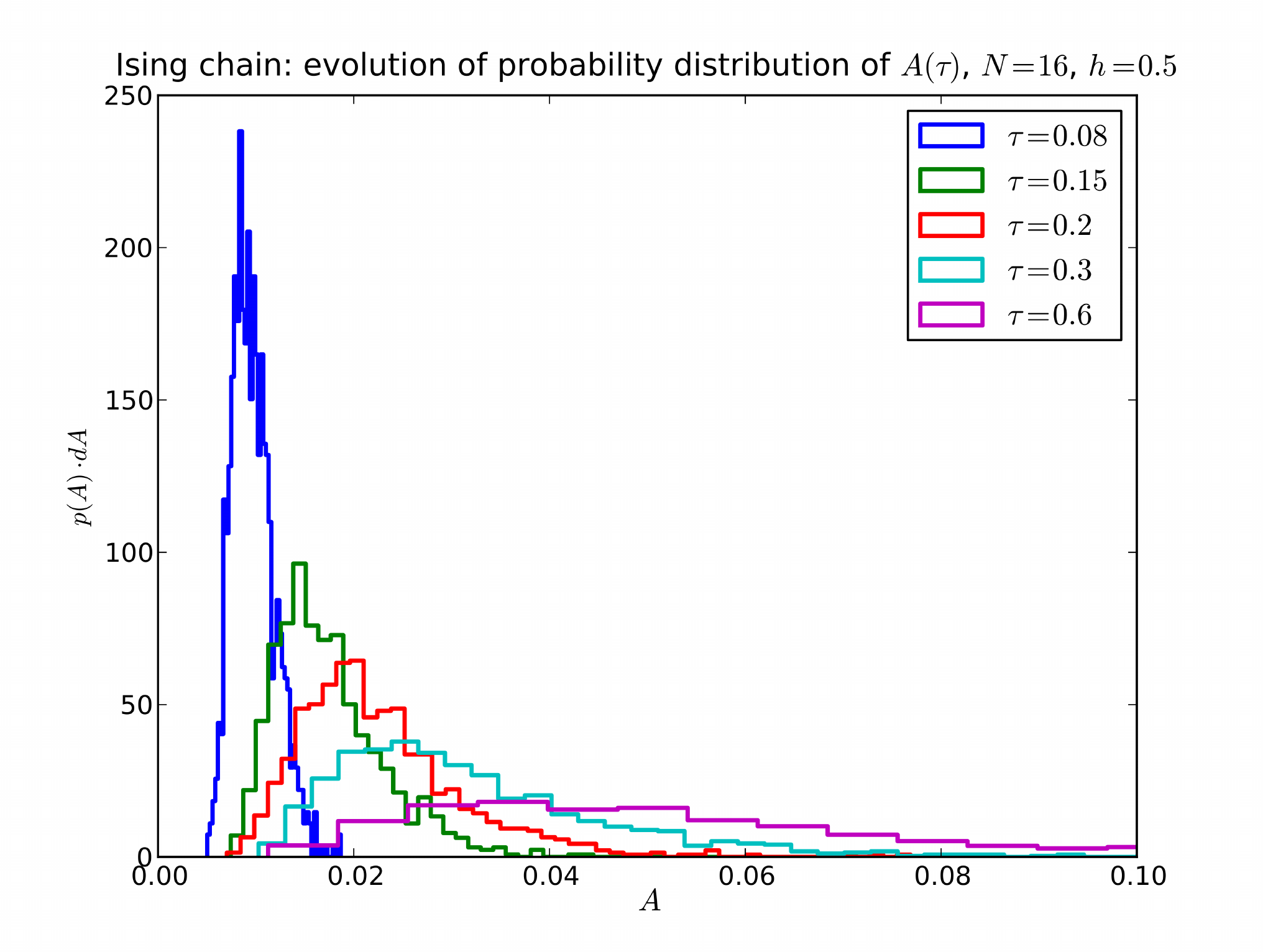}
    \includegraphics[width=0.48\textwidth]{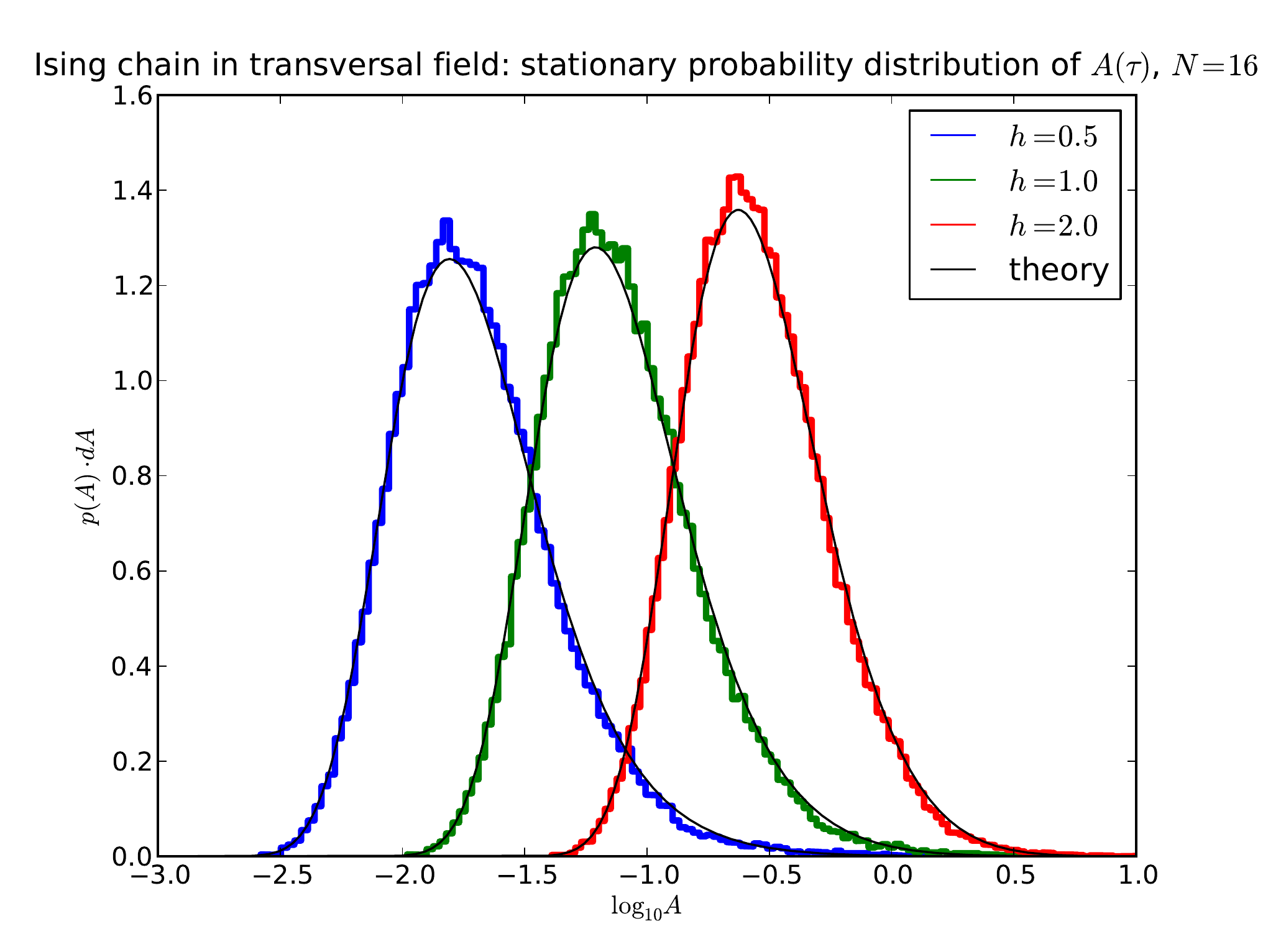}
\end{center}
\caption{ {\bf Upper panel:} The time evolution of the probability distribution of~$A_1(\tau)$. The
histograms are plotted for 1000 realizations of the process with the magnetic field~$h=0.5$.
{\bf Lower panel:} The empirical histograms of~$A_1$ after the distribution becomes stationary (for
large enough~$\tau$, vs. the asymptotic log-normal probability distribution,
plotted for several magnetic fields~$h$. Note that the horizontal axis is plotted in the
$\log$~scale. The histograms shown in the left panel converge for large times~$\tau$ to the ones
in the lower panel.
}
\label{fig:HistogramsAs2}
\end{figure}

Further insight may into the SDE interpretation of the equations may be obtained from
Figs.~\ref{fig:HistogramsAs1}~and~\ref{fig:HistogramsAs2}, where we present numerical results for a 1D Ising
chain of~$N=16$~spins in the transversal magnetic field~$h$, simulated in the imaginary time~$\tau$. 
These figures display typical trajectories of
the local function~$A_1$ (thanks to the translational invariance, we may choose any site), as well
as its probability distribution. Let us note that the stochastic process~$A$ becomes stationary at
large imaginary times, obeying the log-normal distribution.


\end{document}